%% file: remain2.tex
\documentclass[prl,superscriptaddress,twocolumn]{revtex4-2}

\usepackage{amsmath,amsthm,amssymb}
\usepackage[utf8]{inputenc}
\usepackage[american]{babel}
\usepackage{graphicx,xcolor,bbold,titlesec}
\usepackage{braket}
\usepackage{comment}
\usepackage{MnSymbol}

\usepackage[colorlinks,
bookmarksopen,
bookmarksnumbered,
citecolor=red,
linkcolor=red,
pdfstartview=false,
urlcolor=red]{hyperref}
\usepackage{orcidlink}
\usepackage{etoolbox}
\usepackage{tikz}
\usepackage{pgfplots}
\pgfplotsset{compat=1.18}
\usetikzlibrary{calc} 

\def\rhoa{\rho_{\rm wall}}
\def\rhob{\rho_{\rm bulk}}

\def\rr{\textbf{r}}
\def\bb{\textbf{b}}

\usepackage{ifthen}

\newcommand{\andrea}[2][]{%
  \textcolor{orange}{#2}%
  \ifthenelse{\equal{#1}{}}{}{%
    \textcolor{orange}{\textbf{[Andrea: #1]}}%
  }%
}

\def\NA{N_A}

\def\mmu{\boldsymbol{\mu}}
\def\nnu{\boldsymbol{\nu}}
\begin{document}

\def\titleinfo{Quantum Mpemba Effect in Random Circuits}
\title{\titleinfo} 
\author{Xhek Turkeshi~\orcidlink{0000-0003-1093-3771}}
\affiliation{Institut f\"ur Theoretische Physik, Universit\"at zu K\"oln, Zülpicher Strasse 77, 50937 K\"oln, Germany}

\author{
Pasquale Calabrese}
\affiliation{SISSA and INFN Sezione di Trieste, via Bonomea 265, 34136 Trieste, Italy}
\affiliation{International Centre for Theoretical Physics (ICTP), Strada Costiera 11, 34151 Trieste, Italy}

\author{Andrea De Luca~\orcidlink{0000-0003-0272-5083}}
\affiliation{Laboratoire de Physique Th\'eorique et Mod\'elisation,
CY Cergy Paris Universit\'e, CNRS, F-95302 Cergy-Pontoise, France}

\begin{abstract}
The essence of the Mpemba effect is that non-equilibrium systems may relax faster the further they are from their equilibrium configuration. In the quantum realm, this phenomenon arises in the dynamics of closed systems, where it is witnessed by fundamental features such as symmetry and entanglement.
Here, we study the quantum Mpemba effect in charge-preserving random circuits on qudits 
combining extensive numerical simulations and analytical arguments. 
We show that the more asymmetric certain classes  of initial states (tilted ferromagnets) are, the faster they restore symmetry and reach the grand-canonical ensemble.
Conversely, other classes of states (tilted antiferromagnets) do not show the Mpemba effect.
We provide a simple and general mechanism underlying the effect, based on the spreading of nonconserved operators in terms of conserved densities.
Our analysis is based on minimal principles---locality, unitarity, and symmetry. Consequently, our results represent a significant advancement in clarifying the emergence of Mpemba physics in chaotic systems.
\end{abstract}

\date{\today}

\maketitle

The study of far-from-equilibrium dynamics in quantum many-body systems has garnered significant attention, revealing foundational principles across theory and experiments~\cite{rigol2008,polkovnikov2011,fagotti2014,dalessio2016,calabrese2016,essler2016}. 
A key example is the mechanism of thermalization in isolated systems, which explains how chaotic evolution causes the system to look locally as a thermal ensemble~
\cite{deutsch1991,srednicki1994,brenes2021,pappalardi2022,pappalardi2023}. 
Besides, quantum dynamics host a variety of counter-intuitive phenomena. 
A compelling example is the Mpemba effect~\cite{Mpemba}. Initially observed in classical physics~\cite{lasanta2017,lu2017,klich2019,kumar2020,bechhoefer2021,kumar2022,walker2023mpemba,walker2023optimal,bera2023effect} and open quantum systems \cite{nf-19,chatterjee2023multiple,chatterjee2023,kochsiek2022,carollo2021,ias-23,shapira2024,strachan2024nonmarkovian,zhang2024observation,wang2024mpemba,moroder2024thermodynamics}, this effect describes scenarios where the further initial states are from their equilibrium configuration, the faster they relax. 
In isolated systems, the quantum Mpemba effect (QME) has been associated with the anomalous relaxation of a quantum system possessing a conserved charge: the more broken the symmetry is, the faster it may locally reach the grand-canonical equilibrium~\cite{ares2022entanglement}. 
It should be noted that due to its dependence on the initial state, the phenomenon is generally non-universal; yet, its manifestations are ubiquitous, as shown by extensive studies on noninteracting and integrable systems~\cite{ares2022entanglement,ares2023,Murciano_2024,yamashika2024entanglement,caceffo2024entangled,bertini2023dynamics,ferro2024,rylands2023microscopic,chalas2024multiple,ares2024quantum}, 
and in trapped-ion experiments~\cite{joshi2024}.
Remarkably, Ref.~\cite{rylands2023microscopic} provided a first-principle derivation for the QME in integrable systems, offering a level of predictability not present in the classical counterpart.

Beyond the integrable limit, our understanding of the quantum Mpemba effect is much more limited. On the one hand, current numerical results for interacting systems have been limited to a handful of qubits ($N \le 16$), far away from the scaling limit required for thermalization. 
A fundamental problem is how the Mpemba times scale with system size. While this has been addressed in non-interacting and integrable systems, the question remains outstanding for more general, chaotic evolution. At the same time, finding first-principle derivations explanations for chaotic systems is notoriously much more challenging, leaving the mechanism driving the QME a completely open question.

This work sheds light on the emergence of the Mpemba phenomenology in generic many-body quantum systems, focusing on the evolution of entanglement asymmetry as a proxy for the QME.
We employ random unitary circuits (RUCs) with a $U(1)$ global symmetry as minimal models for chaotic unitary dynamics~\cite{nahum2017,nahum2018,khemani2018,keyserlingk2018,fisher2023,nidari2020,sierant2023,richter2023,jonay2024,chan2018,chan20182,chan2019,chan20192,chan2021,Chan2022,rakovszky2019,zhou2019,PhysRevX.8.031058}.
By averaging over the random circuit ensemble, we recast the problem into the computation of a tensor network contraction, allowing us to numerically study large-scale systems that are beyond reach with ordinary quantum simulations. 
We consider circuits whose local Hilbert space contains a spin $1/2$ whose $z$-component is a conserved density. We prepare an initial state that breaks the $U(1)$ symmetry, considering as prototypical examples ferromagnetic and antiferromagnetic states with a finite tilting from the $z$ axis. We study the time evolution of the asymmetry of a given region $A$ of size $N_A$, defined as a measure of the discrepancy of the reduced density matrix from its $U(1)$-symmetrized version~\cite{ares2022entanglement}. 
The relaxation to the Gibbs equilibrium ensures that the asymmetry eventually reaches zero, but the path it takes to reach this equilibrium is the observable of interest in our work. 
Our numerical and analytical analysis provide compelling evidence for the presence of the Mpemba effect in entanglement asymmetry for ferromagnetic states and its absence for antiferromagnetic ones.

We then discuss the fundamental mechanism underlying this phenomenology. First, we show that the asymmetry of a given region  $A$  can be expressed in terms of the expectation values of all non-conserved operators initially supported within  $A$. Recent studies have analyzed how operators grow under Heisenberg evolution, becoming increasingly complex and extending their support. However, more relevant to our discussion are the operators generated by this evolution that consist only of conserved densities.
Initially, the expectation values of non-conserved operators are larger in cases of stronger symmetry breaking, contributing to greater initial asymmetry. Over time, these expectation values decrease as conserved quantities—responsible for driving entanglement entropy production~\cite{rakovszky2019}—are emitted. This perspective demonstrates that larger symmetry breaking amplifies the initial asymmetry but also accelerates entropy production and relaxation, aligning with the Mpemba effect. For ferromagnetic states, this results in faster relaxation, whereas for antiferromagnetic states, entropy remains maximal regardless of the tilting.

This framework, based on the emission of conserved operators, can also be formalized in the limit  $q \to \infty$, where a specific mapping to two coupled simple symmetric exclusion processes (SSEP) becomes feasible. Using macroscopic fluctuation theory (MFT), we provide an analytical treatment of these statistical models, leveraging a hydrodynamic description valid in the large  $N_A$  limit (where  $N_A $ is the size of region  $A$, as defined earlier). However, in this hydrodynamic description, the Mpemba effect is not apparent at small times, as it emerges only at longer times  $t = O(N_A^2)$.

\emph{Model and observables.}
We define our model on a one-dimensional chain of $N$ sites, where 
each site is described by $\mathcal{H}_{2,q}\simeq \mathbb{C}^2 \otimes  \mathbb{C}^q$ of dimension $d=2q$. Conventionally, we refer to the former as a spin-$1/2$ degree of freedom and the latter as a color sector with $q$-orbital states~\footnote{For the sake of presentation, a detailed survay of the technicalities is presented in~\cite{supmat}.}. 
The circuit is composed of two-body independently identically distributed random gates $U_{i,i+1}$ acting on neighboring sites $i,i+1$, and chosen so that $[U_{i,i+1},Z_{i,i+1}]=0$, with $ Z_{i,i+1}\equiv (Z_{i}+Z_{i+1})/2$, with $X_i=\sigma^x_i\otimes I_q$, $Y_i=\sigma^y_i\otimes I_q$, and $Z_i=\sigma^z_i\otimes I_q$ the Pauli matrices at site $i$.
Since the spectrum of $Z_{i,i+1}$ is composed by $S\in \{0,\pm 1 \}$, with degeneracies $d_S = (1+\delta_{S,0}) q^2$, we write $U_{i,i+1} = \bigoplus_S u^{(S)}_{i,i+1}$, where $u_S$ are drawn from the Haar distribution in the unitary group $\mathcal{U}(d_S)$. In this way, by construction, the entire circuit preserves the global charge $Q = \sum_{i=1}^N Z_i/2$. 
Each layer $\tau$ of the circuit follows the brick-wall architecture $U^{(\tau)}=\left(\prod_{i=1}^{N/2}U_{2i-1,2i}\right)\left(\prod_{i=1}^{N/2}U_{2i,2i+1}\right)$~\footnote{Here $U_{N,N+1}\equiv U_{N,1}$ ($U_{N,N+1}\equiv I_1\otimes I_N$) for periodic (open) boundary conditions.}.
Thus, the final circuit for a given depth (also denoted as time) $t$ is given by $U_t = \prod_{\tau=1}^{t} U^{(\tau)}$, and the time evolution is given by $|\Psi_t\rangle = U_t |\Psi_0\rangle$.

\begin{figure*}[t]
    \centering
    \includegraphics[width=\textwidth]{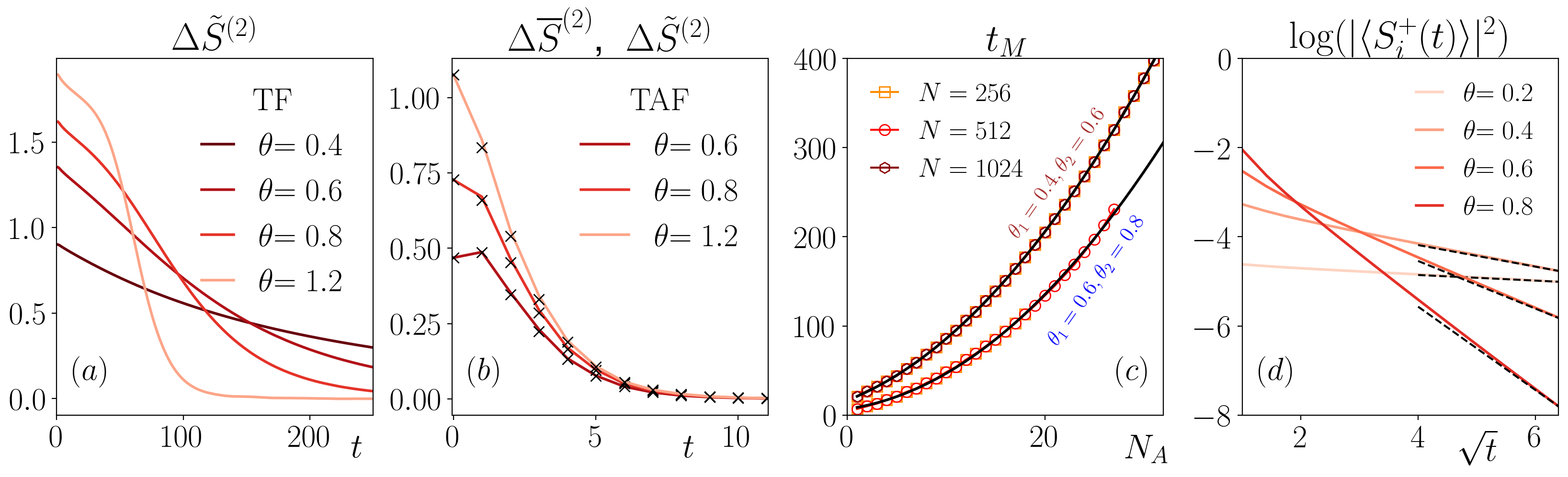}
    \caption{
    Numerical results for spin-$1/2$ circuit ($q=1$).
    (a) Fixing $N=512$ and $NA=16$, we compute the 
    annealed averaged R\'enyi-2 entanglement asymmetry via tensor networks and open boundary conditions starting from the TF state.
    The crossings for different angles are a distinguishing trait of the Mpemba phenomenology. 
    (b) For tilted antiferromagnet (TAF), no crossing, hence no QME, is present. The data is obtained via exact methods for $N=20$ and $N_A=3$ and demonstrates the self-averaging between the quench average (colored lines) and the annealed values (symbols).
    (c) The Mpemba time for different system sizes $N$ as a function of the subsystem size $N_A$. The crossings of $\Delta \tilde S^{(2)}$ are identified comparing curves with different tilting angles, $\theta_1$ and $\theta_2$. The Mpemba time turns out to be a quadratic function of $N_A$. The solid black lines are quadratic fit, highlighting $t_M\sim O(N_A^2)$ as predicted by our analysis.
    (d) Operator spreading of $|\langle S^+_i(t)\rangle|^2$ string average via tensor network contraction, with $|\Psi_0\rangle$ the TF state at different angle $\theta$. We consider $i=N/2$ for a chain of $N=80$ sites. We can approciate the crossing typical of the Mpemba effect. The dashed lines are the prediction obtained from our operator spreading analysis, cf. Eq.~\eqref{eq:magic}, fitting the prefactor of  $\ell_t = \alpha \sqrt{D t}$. 
    } 
    \label{fig:mpembaa}
\end{figure*}

We consider the initial states $\rho_0=|\Psi_0\rangle\langle \Psi_0|$ that are {\it not} symmetric with respect to $Q$, \emph{i.e.}, $[Q,\rho_0]\neq 0$. 
For concreteness, we focus on the tilted ferromagnetic $|\Psi_0^{f}(\theta)\rangle= |\theta^{(0)}\rangle^{\otimes N}$ and antiferromagnetic states $|\Psi_0^{a}(\theta)\rangle= (|\theta^{(0)}\rangle\otimes |\theta^{(1)}\rangle)^{\otimes N/2}$, with $|\theta^{(\alpha)}\rangle = e^{-i Y \theta/2} |\alpha\rangle\otimes |0\rangle$ where $\alpha = 0,1$ refer to the two eigenstates of the $z$ component of the spin and $\ket{0}$ is a reference state in the color basis. We aim to study how the reduced density matrix $\rho_A(t) = \mathrm{tr}_{A_c}(\rho_0)$ of a subsystem $A$ of size $\NA\ll N$ with complement $A_c$ evolves over time. 
Since the sole symmetry of the circuit is the $U(1)$ charge, the dynamics relax $\rho_A(t)$ to the grand-canonical ensemble $\rho_A(\infty)=e^{-\lambda Q_A}/\mathrm{tr}(e^{-\lambda Q_A})$, with $Q_A=\sum_{i\in A} Z_i/2$ the local charge and $\lambda$ fixed by the average initial charge $\mathfrak{q}_0 = N^{-1}\langle \Psi_0|Q|\Psi_0\rangle$. 

As highlighted in~\cite{ares2022entanglement}, since the stationary state is $U(1)$ invariant, symmetry restoration provides a convenient proxy for quantitatively detecting the QME. The parameter $\theta$ is a tunable nod of symmetry breaking, which we quantify via the entanglement asymmetry~\cite{ares2022entanglement}
\begin{equation}
    \Delta S_A^{(n)}(\rho_A) = S_n(\rho_{A,Q})-S_n(\rho_{A})
    \label{eq:asym}
\end{equation}
where $S_n(\rho)\equiv (1-n)^{-1}\ln[\mathrm{tr}(\rho^n)]$ is the Rènyi entanglement entropy, and $\rho_{A,Q}\equiv \sum_{s} \Pi_s \rho_A(t) \Pi_s$ is the charge decohered density matrix (CDDM), with $\Pi_s$ the projector onto the $s$-eigenspace of $Q_A$~\cite{fossati2024entanglement,chen2023,capizzi2023,capizzi2023universal,ares20233,khor2023}. In
Eq.~\eqref{eq:asym}, $ \Delta S_A^{(n)}(\rho_A)\ge 0$, and the equality holds only when $\rho_{A}=\rho_{A,Q}$, \emph{i.e.}, when the state $\rho_A$ is symmetric \cite{Ma2022,Han2023}. 
We say $\rho_{A}$ less symmetric than $\sigma_{A}$ if $ \Delta S_A^{(n)}(\rho_A)> \Delta S_A^{(n)}(\sigma_A)$, and accordingly the former is further from equilibrium than the latter. 
Thus, the quantum Mpemba effect manifests when (i) $\Delta S_A^{(n)}(\rho_A(0))> \Delta S_A^{(n)}(\sigma_A(0))$ and (ii)  $\Delta S_A^{(n)}(\rho_A(\tau))< \Delta S_A^{(n)}(\sigma_A(\tau))$ for $\tau > t_M$, with $t_M$ is the Mpemba time, cf.~\cite{rylands2023microscopic}. 

\emph{Numerical methods.} 
Our analysis combines extensive numerical computations and analytical methods, which we briefly review below, see also End Matter and Ref.~\cite{supmat}.
First, we remark that the problem of interest is stochastic due to the randomness of the gates. 
Thus, we will focus on average quantities over the circuit realizations. 
Specifically, denoting 
by $\mathbb{E}(\bullet) $ 
the circuit average, we will study the quenched and the annealed averages of the entanglement asymmetry, given respectively, by
\begin{equation}
        {\Delta \overline{S}_A^{(n)}} \equiv \mathbb{E}\left[\Delta S_A^{(n)}(\rho_A)\right], \quad
        {\Delta \tilde{S}_A^{(n)}} \equiv \frac{1}{1-n}\ln\left[ \frac{\mathcal{P}_{A,Q}^{(n)}}{\mathcal{P}_{A}^{(n)}}\right],
    \label{eq:averages}
\end{equation}
where $\mathcal{P}_A^{(n)}= \mathbb{E}[\mathrm{tr}(\rho_A^n)]$ and $\mathcal{P}_{A,Q}^{(n)} = \mathbb{E}[\mathrm{tr}(\rho_{A,Q}^n)]$. 
Both quantities in~\eqref{eq:averages} are computable at any Rènyi order $n$ using exact numerical computations. 
Despite the limited system sizes attainable with exact methods, our results highlight strong self-averaging features of the entanglement asymmetry, with ${\Delta \overline{S}_A^{(n)}(t)} = {\Delta \tilde{S}_A^{(n)}}(t) +O(e^{-\gamma N})$. 
This fact enables us to obtain quantitative predictions on quench averages from the quantities $\mathcal{P}_A^{(n)}$ and $\mathcal{P}_{A,Q}^{(n)}$, both expressible as tensor network contractions. 
This will allow extensive numerical simulations for $\Delta\tilde{S}_A^{(n)}$ and analytical insights.

\emph{Quantum Mpemba effect for qubits. }
We begin our discussion showcasing the numerical analysis at $q=1$, summarized in the Fig.~\ref{fig:mpembaa}. 
We consider the tilted ferromagnetic (TF) state $|\Psi_0^f(\theta)\rangle$ for $N=512$ and $N_A=16$, we find that R\'enyi 2 entanglement asymmetry obtained via tensor networks (cf. Fig.~\ref{fig:mpembaa}(a)) exhibit the typical crossing pattern of the QME~\cite{ares2022entanglement}: 
the more the $U(1)$ symmetry is initially broken, the faster it is restored. 

The situation drastically differs for tilted antiferromagnetic (TAF) initial  state $|\Psi_0^a(\theta)\rangle$, cf. Fig.~\ref{fig:mpembaa}(b). Here, already for the small systems accessible via exact diagonalization $N=20$ and $N_A=3$, the crossing, hence the Mpemba effect, is absent. 
In the same figure, we also present the strong self-averaging signature of the circuit, comparing the quench averages (colored lines) with the annealed averages (crosses). 

From the tensor network simulations, we extract the Mpemba time $t_M$ as the crossing point between two curves with different $\theta_1<\theta_2$ angles for initial TF states.  In Fig.~\ref{fig:mpembaa}(c) we highlight the empirical observation that $t_M\sim O(N_A^2)$ for system sizes up to $N\le 1024$, which we will corroborate below with analytical arguments.

\emph{QME and operator spreading.} 
To elucidate the mechanism behind the observed phenomenology, it is useful to clarify the connection between the asymmetry and operator spreading~\cite{khemani2018,keyserlingk2018} and for simplicity we look at $n = 2$ in \eqref{eq:asym} and set momentarily $q=1$. 
We introduce a full orthonormal basis of operators with support in $A$ as $B^{\mmu}=\prod_{i\in A} B_i^{\mu_i}$, where $B_i^{\mu_i} \in \{\mathbb{1}_i, S^{\pm}_i := (X_i \pm i Y_i)/{\sqrt{2}},\sigma^z_i\}$ and with $\mathcal{S}(A)$ the set of all $\boldsymbol{\mu} = (\mu_1,\ldots, \mu_{N_A})$. In this way, we can express
$\rho_{A,t} = \sum_{\vec{\mu}\in \mathcal{S}(A)} \langle B^{\boldsymbol{\mu}\dagger}\rangle_t B^{\boldsymbol{\mu}}$. Restricting the sum over strings commuting with $Q_A$, we can express
$\rho_{A,Q,t} = \sum_{\vec{\mu}\in \mathcal{S}(A), [B^{\boldsymbol{\mu}},Q_A]=0} \langle B^{\boldsymbol{\mu}\dagger}\rangle_t B^{\boldsymbol{\mu}}$. 
It follows that, when computing the asymmetry, strings commuting with 
$Q_A$ cancel out, and for small $\theta$ we have
\begin{equation}
\label{eq:asynoncons}
    {\Delta \tilde{S}_A^{(2)}} \simeq  
\frac{
\sum_{{\mmu \in \mathcal{S}(A), [B^{\mmu}, Q_A] \neq 0}} |\langle B^{\mmu}\rangle_t|^2
}{\sum_{\mmu \in \mathcal{S}(A)} |\langle B^{\mmu}\rangle_t|^2}
\;,
\end{equation}
where in the numerator the sum is restricted to off-diagonal strings. To further analyse this expression, we decompose the Heisenberg evolved $B^{\mmu}(t) = \sum_{\nnu} c^{\mmu}_{\nnu}(t) B^{\nnu}$. The squared weights $|c^{\mmu}_{\nnu}(t)|^2$  
set the support of the operator that expands ballistically, similarly to what occurs in random circuits without conserved charges and this is the relevant feature for out-of-time order correlators~\cite{khemani2018, PhysRevX.8.031058, keyserlingk2018, nahum2018}. In contrast, Eq.~\eqref{eq:asynoncons} depends on the coherent sums within the expansion of $B^{\mmu}(t)$. For this, it is crucial to observe that the conserved densities exhibit diffusive evolution~\cite{khemani2018, rakovszky2019, PhysRevX.8.031058}
\begin{equation}
\label{eq:diffspreadZ}
    Z_i(t) = \sum_{j} K_{i-j}(t) Z_j  + \tilde{Z}_i(t)
\end{equation}
where $K_{j}(t)$ is a lattice diffusion Kernel approximated in the continuum by  $K_j(t) \sim \exp(-\frac{j^2}{2 D t})/\sqrt{2 \pi D t}$, with
$D$ the diffusion constant. The operator $\tilde{Z}_i(t)$ contains all the other strings produced by time evolution. In particular, under circuit average $\mathbb{E}(\tilde{Z}_i(t)) = 0$. So, each operator $Z_i$, once generated, spreads diffusively and coherently, so that its expectation value rapidly converges to $\langle Z_i \rangle_t \to 2\mathfrak{q}_0$. Thus, the denominator of Eq.~\eqref{eq:asynoncons} is dominated by the strings $B^{\mmu} = Z_{i_1} \ldots Z_{i_m}$ of local densities  which move away from each other diffusely, leading to $e^{-S_2(\rho_{A,t})} = 2^{-\NA} \sum_{\mmu \in \mathcal{S}(A)} | \langle B_t^{\mmu} \rangle|^2 \stackrel{Dt\gg N_A^2}{\longrightarrow} e^{-N_A s_2(\mathtt{q}_0)}$, with $D$ the diffusion constant and $s_2(\mathfrak{q}_0) = \log(2) -\log(1 + 4 \mathfrak{q}_0^2)$ the equilibrium second-Renyi density.

These terms are not present in the numerator of Eq.~\eqref{eq:asynoncons}, which in fact never saturates but continues to decrease. To see it, we first associate with each string $\mmu$ a \textit{raising charge}, assigning $\pm 1$ to $S^{\pm}_i$ and zero otherwise, finally adding the contribution of each site $i$. In this way, the off-diagonal strings are those characterized by a non-zero raising charge. Moreover, the $U(1)$ symmetry dictates that this raising charge is conserved by the Heisenberg evolution. Taking the expectation value on the tilted (anti)ferromagnetic states $\ket{\Psi_0}$, each local element $S^{\pm}_i$ contributes with a factor $O(\theta)$, so for small $\theta \ll N_A^{-1/2}$, the relevant contribution comes from operators $B^{\mmu} = S^{\pm}_i$ with $i \in A$, having minimum non-zero charge $\pm1$. As the two signs give equal contributions, we focus on the plus sign. During the evolution of Heisenberg $S^{+}_i(t)$ will produce many more strings but focusing on small $\theta$, we can exclude strings containing other ladder operators. Thus the relevant contribution comes from the strings $B^{\nnu} = S^{+}_j Z_{i_1} \ldots Z_{i_m}$. In this description, we can then imagine the dynamics of the $S^{+}_i$ operator as characterized by a random walk
$S^+_i \to S^+_j$
that moves the raising charge, associated with the emission of  local $Z_i$ densities.
To compute  $\langle S_j^+(t) \rangle = \braket{\Psi_0 | S^+_i(t) |\Psi_0}$, we need to estimate the number of $Z_i$'s operators that are generated by the Heisenberg dynamics of $S^+_i$ over time $t$. Since each operator diffuses, the number of emitted operators can at most fill a region of width $\ell_t = O(\sqrt{D t})$ around site $j$. Once the $Z_i$ operators produced within this region have spread out, their expectation value converges to the average magnetization, a mechanism analogous to the saturation of the second Renyi entropy in the numerator. Thus,
we obtain
\begin{equation}
|\langle S_+(t)\rangle|^2\sim\mathbb{E} [|\langle\Psi_0 |S^+_i(t) |\Psi_0\rangle|^2]\sim
\frac{\theta^2}{2} e^{-\ell_t s_2(\mathtt{q}_0)}\;,
    \label{eq:magic}
\end{equation}
and, accounting for all $i \in A$, we have
\begin{equation}
\label{eq:asymmetry}
\Delta \tilde S_A^{(2)}
\sim N_A \theta^2 e^{(N_A - \ell_t) s_2(\mathtt{q}_0)}
\end{equation}
The details of the length scale $\ell_t \sim \alpha \sqrt{D t}$ depend on the intricate interplay between productions of $Z_i$ operators and their spreading, but its diffusive scaling is robust. For the
ferromagnetic state $s_2(\mathtt{q}_0^f) \sim \theta^2 /2$ and Eq.~\eqref{eq:asymmetry} predicts the presence of the QME due to the fact that a higher $\theta$ produces greater initial asymmetry but also faster relaxation. We also estimate the Mpemba time as $t_M \propto N_A^2/D $. Instead, for antiferromagnetic states $s_2(\mathtt{q}_0^a) = \log 2$ and the effect disappears.

Our numerical analysis demonstrate the validity of Eq.~\eqref{eq:magic}, as appreciated in Fig.~\ref{fig:mpembaa}(d) where we evaluate $\mathbb{E} [|\langle\Psi_0 |S^+_i(t) |\Psi_0\rangle|^2]$ in the middle of the chain $i = N/2$.

If $q>1$, the mechanism of operator spreading is modified by the appearance of strings of operators acting in the color sector. As discussed in~\cite{nahum2018, 
keyserlingk2018, PhysRevX.8.031058, khemani2018}, these operators spread ballistically and incoherently, with a speed equating the entanglement velocity $v_E$~(see~End Matter).
In particular $v_E$ remains finite even for $\theta \to 0$, and this contribution dominates the decay of the asymmetry at large times $t$. So, in the presence of extra degrees of freedom the QME becomes a subleading correction which could be hard to be seen in practice as the asymmetry can be already rather small when the crossing occurs.

\emph{Large-$q$ hydrodynamic description.} 
The scenario above can be corroborated in the limit of $q$ large. By averaging over two (or more) replicas of the circuit, this limit has allowed the formalization of the so-called membrane picture: it associates the growth of entanglement in time with the elastic energy of an interface in the $(x,t)$ plane, pinned on one end to the edge of the $A$ region. In the presence of $U(1)$ symmetry, this picture is modified. Although at $q \to \infty$, as we explained above, the non-conserved degrees of freedom are dominant, one can formally look at the residual dynamics in the spin sector~\cite{rakovszky2019,supmat}. This approach has been used to show that the membrane emits $Z_i$'s operators in each replica that diffuse independently (see Fig.~\ref{fig:sketchmembrane}). Identifying these operators as two species of particles, one obtains the effective dynamics of two SSEPs, coupled in proximity of the membrane. Calculating the purity of the CDDM can be done in a similar way by taking into account that the projector on a given sector of the magnetization actually produces an initial density of particles within the region $A$~\cite{supmat}. 
\begin{figure}
    \centering
    \begin{tikzpicture}
\input{sketchmembrane}
\end{tikzpicture}
    \caption{A sketch of the membrane picture in the presence of $U(1)$ symmetry. The interface is pinned at the edge of the interval $A$ and performs a random walk while it emits $Z_i$'s densities in both replicas (blue and red). Saturation of the entropy is due to the membrane exiting the left boundary of the system, so that the emission stops.}
    \label{fig:sketchmembrane}
\end{figure}
However, this derivation is formally restricted
to the regime of $t\ll N_A$, before the colour degrees of freedom (which dominate at large q) force the membrane off the edge of the system causing entropy saturation. For this reason, this approach does not allow the Mpemba effect to be observed directly. In the regime of short times and large $N_A$, a hydrodynamic description via macroscopic fluctuation theory (MFT)~\cite{Derrida2009,Saha_2023} is possible. We can express $\mathcal{P}_{A,Q}^{(2)}$ as a stationary phase integral, namely
\begin{equation}
   \mathcal{P}_{A,Q}^{(2)} \simeq \mathcal{P}_A^{(2)}\int \frac{dk}{2\pi} e^{-N_A f_t(k,\theta)},
\end{equation}
with $f_t(k,\theta)=k^2 [\sin^2(\theta)/4-\sqrt{t}g(\theta)/N_A]$ the saddle point function, and $g(\theta)=\theta^2/(2\sqrt{\pi}) + (6\sqrt{2}-5)\theta^4/(12\sqrt{\pi}) + O(\theta^6)$ known perturbatively for small $\theta$. This leads to 
\begin{equation}
     \Delta\tilde{S}_A^{(2)}(t)\simeq \Delta\tilde{S}_A^{(2)}(0) -\frac{2 \sqrt{t} g(\theta ) }{\sqrt{\pi N_A^3} \sin(\theta)^3},
\end{equation}
with $\Delta\tilde{S}_A^{(2)}(0)$ the initial state asymmetry and $g(\theta)/\sin^3(\theta)$ a monotonously decreasing function of $\theta$, confirming that Mpemba physics is not visible at short times $t\ll N_A$.

\emph{Conclusion.}
In this paper, we investigated the quantum Mpemba effect in random unitary circuits with a $U(1)$ conservation law. 
Our approach combines exact numerical methods and tensor network simulations, with analytical results via operator spreading and macroscopic fluctuation theory.
For qubits, we demonstrate the presence (absence) of the QME for tilted ferromagnetic (antiferromagnetic) initial states, depending on the state's equilibrium second-R\'enyi density. 
The operator spreading picture allow us to pinpoint the Mpemba time as scaling quadratically with $N_A$, in agreement with our numerical results. 
As the argument readily extends to generic chaotic Hamiltonian~\cite{rakovszky2019}, we expect our results to apply to generic many-body systems.



We envision multiple follow-ups. 
While ideal unitary evolution is closed, errors and measurements are crucial in the noisy-intermediate scale quantum device era. It would be, therefore, interesting to study if the Mpemba phenomenology survives when including such non-unitary generators~\cite{fisher2023,Potter2022}, for instance, the interplay with projective~\cite{skinner2019,sierant2022,sierant20222,agrawal2022,barratt2022,oshima2023,klocke2023} and weak monitoring~\cite{cao2019,alberton2021,buchhold2021,fava2023,poboiko2023,muller2022,loio2023,poboiko2023measurementinduced,chahine2023entanglement,turkeshi2022e,turkeshi2024density,legal2023,gal2024entanglement,leung2023theory}.
Another question of interest is whether Mpemba phenomenology arises beyond the standard paradigm of thermalization, namely in deep thermalization. It would be interesting to explore the relationship between deep thermalization timescales and Mpemba times in the presence of symmetry and conserved quantities~\cite{ippo}. We leave these investigations open for future work. 

\begin{acknowledgments}
{\em Acknowledgments.}
We thank C. von Keyserlingk for insightful discussions. 
X.T. acknowledges support from DFG under Germany's Excellence Strategy – Cluster of Excellence Matter and Light for Quantum Computing (ML4Q) EXC 2004/1 – 390534769, and DFG Collaborative Research Center (CRC) 183 Project No. 277101999 - project B01. 
He also thanks the Institut Henri Poincaré (UAR 839 CNRS-Sorbonne Université) and the LabEx CARMIN (ANR-10-LABX-59-01) for their support during the completion stage of this work. 
PC acknowledges support from ERC under Consolidator grant number 771536 (NEMO). ADL acknowledges support
by the ANR JCJC grant ANR-21-CE47-0003 (TamEnt).

{\em Data Availability.} 
Numerical data and codes for the manuscript are publicly available in Ref.~\cite{dataavail}.

\emph{Note added.} During the completion of this manuscript, a complementary work appeared in the arXiv~\cite{liu2024symmetry}. When overlapping, our results are in agreement. 
\end{acknowledgments}

\onecolumngrid 
\newpage 

\appendix
\section{End Matter}

\subsection{Replica Tensor Network}
We present details of the computational methods employed for $n=2$ replica~\footnote{The replica limit $n\to 1$ is complicated by a non-trivial interplay between hydrodynamic and non-hydrodynamic modes~\cite{keyserlingk2018}.}.
Given the system state for a realization $\rho = U_t |\Psi_0\rangle\langle \Psi_0| U_t^\dagger$, the purity of the reduced density matrix on $A$ is given by $P_A = \mathrm{tr}\left( \mathcal{F}_A[\rho\otimes \rho]\right) $, with $\mathcal{F}_A = \bigotimes_{i \in A} \mathcal{F}_i$, and the swap operators $\mathcal{F}(|x\rangle \otimes |y\rangle) = |y\rangle\otimes |x\rangle$, with $\{|x\rangle\}$ a basis of $\mathcal{H}_{2,q}$. 
It is convenient to pass to the superoperator formalism, with $\rho \mapsto |\rho\rrangle$, and $U_t \rho_0 U_t^\dagger \mapsto (U_t \otimes U_t^*) |\rho_0\rrangle$ so that ${P}_A = \llangle \mathcal{F}_A| U_t \otimes U_t^* \otimes U_t \otimes U_t^* |\rho_0^{\otimes 2} \rrangle$. Hence, by linearity and the fact that  $U_{i,i+1}^{(\tau)}$ are i.i.d., the average purity $\mathcal{P}_A^{(2)} = \mathbb{E}(P_A)$ amounts to computing second moments over the Haar ensemble for each gate, namely $\mathcal{T}_{i,i+1}=\mathbb{E}_\mathrm{Haar}\left[(U_{i,i+1} \otimes U_{i,i+1}^*)^{\otimes 2}\right]$.
Its explicit expression for generic $q$ can be obtained using the Weingarten calculus~\cite{supmat,Collins2006}. 
For instance, for $q\ge 2$ we have
\begin{equation} 
\begin{split}
   \mathcal{T}_{i,i+1} = \sum_{\mu,\nu=\pm} \sum_{\{r=\pm 1/2\}} \sum_{\{b=\pm 1/2\}} \mathcal{W}_{\mu,r_1,b_1,r_2,b_2}^{\nu,r_1',b_1',r_2',r_2'} |\mu,r_1,b_1\rrangle_i|\mu,r_2,b_2\rrangle_{i+1}\llangle \nu,r_1',b_1'|_i\llangle \nu,r_2',b_2'|_{i+1}\;,
\end{split}
    \label{eq:transf}
\end{equation}
where $|\mu,r,b\rrangle_i$ are 8-dimensional vectors on site $i$-th and defined the coefficients~\cite{rakovszky2019}
\begin{equation}
\begin{split}
\mathcal{W}_{\mu,r_1,b_1,r_2,b_2}^{\nu,r_1',b_1',r_2',r_2'}= \sum_{S_1, S_2}\frac{\mathrm{w}_{\mu\nu}(S_1,S_2)}{d_{S_1}d_{S_2}-\delta_{S_1,S_2}} \delta_{r_1+r_2 ,S_1}\delta_{b_1+b_2 ,S_2}\delta_{r_1'+r_2' ,S_1}\delta_{b_1'+b_2',S_2}\;
\end{split}
\end{equation}
and $\mathrm{w}_{+}(S_1,S_2)=1$ and $\mathrm{w}_{-}(S_1,S_2)=-\delta_{S_1,S_2}/d_{S_1}$. 
Therefore, averaging over the Haar ensemble simplifies the relevant replica local Hilbert space from $(2q)^8$ to the $64$-dimensional space generated by the on-site replica basis $\mathcal{B}_q\equiv \{|\mu,r,b\rrangle_i\}$, valid for any $q\ge 2$ and where each $|\mu,r,b\rrangle_i$ is a $(2q)^8$-dimensional vector. 
A similar expression for $\mathcal{T}_{i,i+1}$ can be obtained for $q=1$, which, in this limit, is a $36\times 36$ matrix because of degenerate basis vectors in $\mathcal{B}_q$, see~\cite{supmat} for details. 
In this basis, the initial state reads $| \rho_0^{\otimes 2} \rrangle =\bigotimes_{i=1}^N | \Theta^{(i)}\rrangle$, with $| \Theta^{(i)}\rrangle$ depending on the tilted ferromagnetic and antiferromagnetic initial value, and the average initial charge $\mathfrak{q}_0$. For instance, for the tilted ferromagnet, we have
\begin{equation}
    | \Theta^{(i)}\rrangle = \sum_{\mu=\pm }
    \sum_{r,b=\pm1/2} [\cos(\theta)]^{r+b+1}|\mu,r,b\rrangle\;.
\end{equation}
On the other hand, the final state for the purity is $|\mathcal{F}_A\rrangle  = \bigotimes_{i=1}^N | F^{(i)}\rrangle$ with $| F^{(i)}\rrangle = |-,-\frac{1}{2},-\frac{1}{2}\rrangle$ for $i\in A$ and $| F^{(i)}\rrangle=|+,-\frac{1}{2},-\frac{1}{2}\rrangle$ otherwise. 
Collecting all these facts, we arrive at an effective statistical mechanics model for the purity~\cite{keyserlingk2018}. Each layer of the original circuit maps, upon replica trick and Haar average, to the same transfer matrix $\mathcal{T}=\left(\prod_{i=1}^{N/2}\mathcal{T}_{2i-1,2i}\right)\left(\prod_{i=1}^{N/2}\mathcal{T}_{2i,2i+1}\right)$, hence $\mathcal{P}_A^{(2)} = \llangle  \mathcal{F}_A| \mathcal{T}^t|\rho_0^{\otimes 2}\rrangle$. 
As anticipated, this fact has an operational consequence: $\mathcal{T}$ is expressible as a matrix product operator, while both initial and final states are matrix product states, and $\mathcal{P}_A^{(2)}$ amount to computing contractions~\cite{supmat}.

While we discussed mainly the purity $\mathcal{P}_A^{(2)}$, the statistical mapping can be easily extended to $\mathcal{P}_{A,Q}^{(2)}$. The only difference is in the initial state, which requires accounting for the decoherence channel $\mathcal{C}_A(\bullet) = \sum_{s} \Pi_s(\bullet)\Pi_s$. Using discrete Fourier analysis, we have 
\begin{equation}
    \mathcal{C}_A(\bullet) =  \frac{1}{\NA+1}\sum_{k \in \mathcal{K}} e^{- i \pi k Q_A} (\bullet) e^{+ i \pi k Q_A}\;,
    \label{eq:dft}
\end{equation}
with $\mathcal{K}=\{ 2\pi j/(N_A+1) \;|\; j=0,1,\dots,N_A\}$ the moments. Eq.~\eqref{eq:dft} allows us to express the decohered purity as $\mathcal{P}_{A,Q}^{(2)}=\sum_{k\in\mathcal{K}} \llangle  \mathcal{F}_{A,k}| \mathcal{T}^t|\rho_0^{\otimes 2}\rrangle/(2N_A)$, again efficiently implementable in tensor-network computations,  
with each moment $k$ fixing the initial replica state $| \mathcal{F}_{A,k}\rrangle\equiv \bigotimes_{i=1}^N | F^{(i)}(k)\rrangle$, with $| F^{(i)}\rrangle=|+,-\frac12,-\frac12\rrangle$ if $i\in A_c$ and, otherwise~\eqref{eq:dft} leads to
\begin{equation}
\begin{split}
    &| F^{(i)}(k)\rrangle = \sum_{r,b=\pm1/2} i^{r-b}\frac{[\sin(k)]^{r+b+1}}{[\cos(k)]^{r+b-1}}|-,r,b\rrangle\;.
    \label{eq:fourier}
\end{split}
\end{equation}
Summarizing, the annealed average entanglement asymmetry $\Delta \tilde{S}_A^{(2)}$ maps to a difference in free energies, respectively $-\ln[\mathcal{P}_A^{(2)}]$ and $-\ln[\mathcal{P}_{A,Q}^{(2)}]$, both computable via tensor network methods~\cite{itensor,Schollw_ck_2011,keyserlingk2018,turkeshi2024hilbert} for system sizes larger than exact computation techniques permit. 

\begin{figure}[ht]
    \centering    \includegraphics[width=0.6\columnwidth]{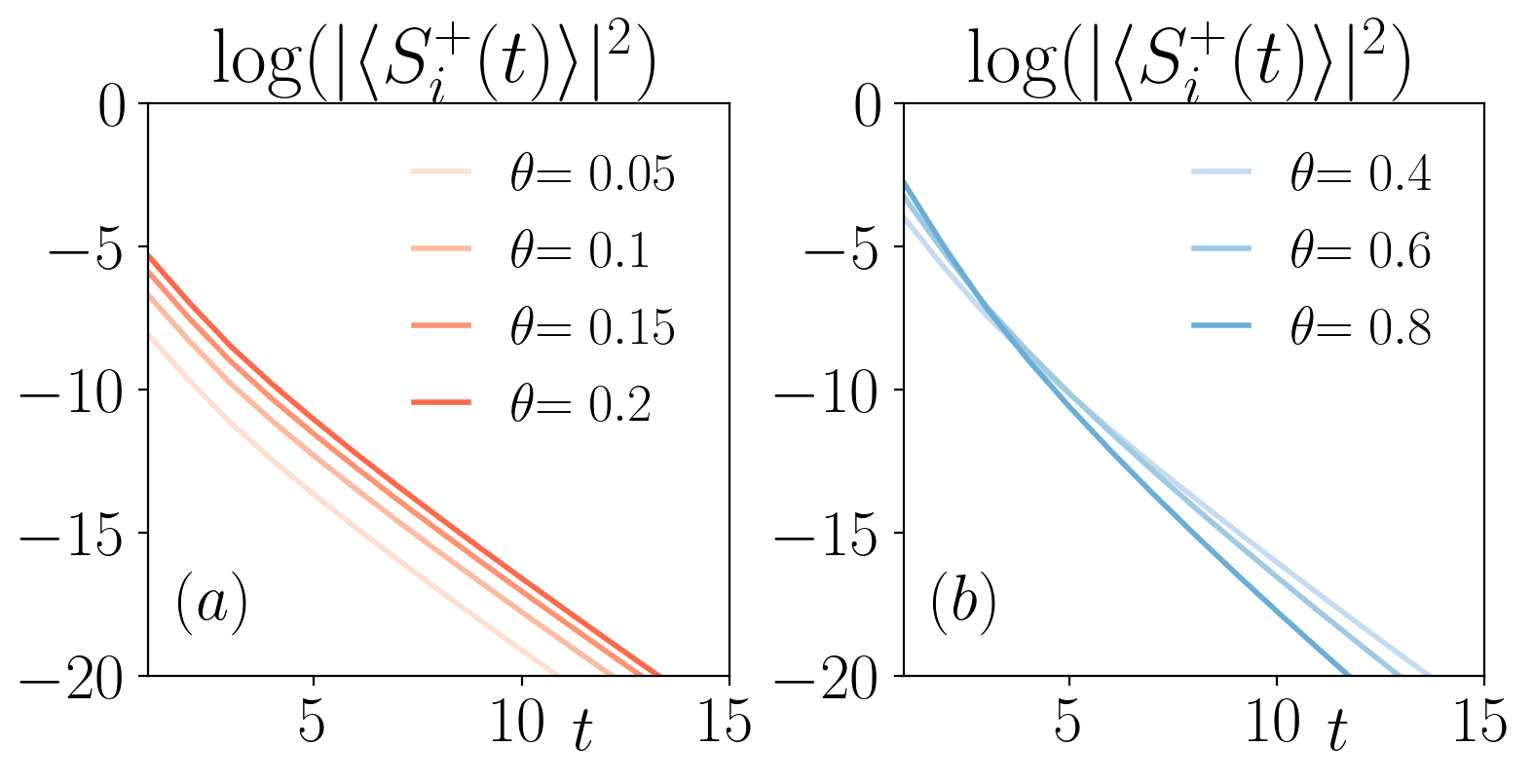}
    \caption{Operator spreading for  $N = 80$  with  $i = 40$  of a single  $S^+_i$  operator is analyzed. (a) At small values of  $\theta$ , as predicted by our analysis, the leading behavior is proportional to  $\exp(-\gamma t)$, regardless of the angle  $\theta$. The time dependence remains consistent, differing only by a shift due to the initial state constant. (b) For larger  $\theta$, a crossover effect emerges between the ballistic behavior ( $\gamma t$ ) and the diffusive behavior ( $\sqrt{D t}$ ). 
    }
    \label{fig:r3}
\end{figure}

\subsection{Operator spreading for $q>1$ \label{sec:opspreadqgt1}}
We complement the discussion on the Main Text on the operator spreading for the case $q>1$. Following~\cite{khemani2018, keyserlingk2018}, we introduce a suitable basis of operators in the $\mathbb{C}^2 \otimes \mathbb{C}^q$ space
\begin{equation}
    B_i^{\mu} = \{ \sigma^{\mu_s}_i \mathcal{X}_i^{\mu_q^{(1)}} \mathcal{Z}_i^{\mu_q^{(2)}} \}_{\mu} \;.
\end{equation}
where $\mu_s$ describe the spin part as in the main text.
Following~\cite{keyserlingk2018}, we denote generalized Pauli operators acting on the color sectors of the $i$-th site as $\mathcal{X}_i\equiv \mathbb{1}_2\otimes J^x_i$ and $\mathcal{Z}_i\equiv \mathbb{1}_2\otimes J^z_i$, with $J^x = \sum_{a=0}^{q-1}|a\rangle\langle (a+1)\mod q|$ and $J^z = \sum_{a=0}^{q-1} e^{-i 2\pi a/q} |a\rangle\langle a|$, with $\mu_q^{(1,2)} \in \{0, \ldots, q-1\}$. 
Overall, the operators $B_i^\mu$ are traceless and orthogonal under the Frobenius inner product $    \frac{1}{2q}\mathrm{tr}[(B_i^{\mu})^\dag B_j^{\nu}] = \delta_{ij} \delta^{\mu\nu}$.
The states considered in the main text are written as product states of the local $|\theta^{(\alpha)}\rangle = e^{-i Y \theta/2} |\alpha\rangle\otimes |0\rangle$ and we choose the reference state $\ket{0}$ for the color degrees of freedom on each site as the eigenstate of $\mathcal{Z}_j$: $\mathcal{Z}_j \ket{0}_j = \ket{0}_j$. This automatically implies that $\braket{0 | \mathcal{X}_j^k | 0}=0$ for all $k = 1,\ldots, q-1$. As usual,
we define a basis of operators in the whole system as
$B^{\mmu} = \prod_{i} B^{\mu_i}_i$. As a paradigmatic example, we focus once again on the spreading of a single-site off-diagonal operator in the middle of the chain $S^+_{i=N/2}(t) = \sum_{\nnu} c_{\nnu}(t) B^{\nnu}$. In this case, we can argue as in the main text, about the fact that in the limit of small tilting $\theta$, the leading contribution comes from operators containing a certain number of density operators $Z_i$'s. However, these operators are flanked by an operator in the color sector, which spreads ballistically and incoherently, as it happens for circuit without conserved quantities. In these cases, one can introduce a propagator
\begin{equation}
    R[y; t] = \sum_{\nnu, \mathcal{E}(\nnu) = y} |c_{\nnu}(t)|^2, 
\end{equation}
where the sum is restricted to the strings ending on the right in $y$. A similar quantity can be defined for the spreading of the left end but we neglect it here. In random circuit, $R[y; t]$ is known to be associated with the probability distribution of a drifted Brownian motion, moving at the so-called butterfly velocity $v_B>0$ while spreading diffusively. More generally, one expects a large deviation principle~\cite{PhysRevB.98.144304} with $R[v t; t] = e^{-t I(v)}$, where $I(v)$ is the rate function (also known as the velocity dependent Lyapunov exponent) satisfying $v_B = \operatorname{argmin}[I(v)]$. However, because of \eqref{eq:expectcolor}, not all strings $\nnu$ can contribute but only a fraction $1/q$ (per site) of them. This leads to the estimate
\begin{equation}
    |\langle S^+_{N/2}(t) \rangle|^2 \sim \sum_y e^{-t (I(y/t)} q^{-y} \sim e^{-v_E t} \;, \quad v_E = \min_{v} [I(v) + \log(q)] \;.
\end{equation}
So, we expect that for $q>1$, the leading decay of the squared expectation value of off-diagonal operators decays linearly in $t$, with a rate related to the entanglement velocity. Note in particular that $v_E$ remains finite even when $\theta \to 0$. We expect the $\theta$-dependent decay of the asymmmetry responsible for QME discussed in Eq.~\eqref{eq:asymmetry} to remain as a subleading correction. Thus, while the QME is still expected, its visibility might be more problematic in practice in the presence of extra degrees of freedom, see Fig.~\ref{fig:r3}.

\bibliography{newbib}
\bibliographystyle{apsrev4-2}

\onecolumngrid
\newpage

\setcounter{secnumdepth}{2}
\setcounter{equation}{0}
\setcounter{figure}{0}
\renewcommand{\thetable}{S\arabic{table}}
\renewcommand{\theequation}{S\arabic{equation}}
\renewcommand{\thefigure}{S\arabic{figure}}
\titleformat{\section}[hang]{\normalfont\bfseries}
{\thesection:}{0.5em}{\centering}
\clearpage
\begin{center}
\textbf{\large Supplemental Material}
\end{center}
\setcounter{equation}{0}
\setcounter{figure}{0}
\setcounter{table}{0}

\renewcommand{\theequation}{S\arabic{equation}}
\renewcommand{\thefigure}{S\arabic{figure}}
\renewcommand{\bibnumfmt}[1]{[S#1]}
\newtheorem{thmS}{Theorem S\ignorespaces}

\newtheorem{claimS}{Claim S\ignorespaces}

In this Supplemental Material we:
\begin{itemize}
    \item Present additional details of our setup and on the numerical implementation;
    \item Present additional discussions on the $q\to\infty$ limit.
    \item Present the detailed analysis of the operator spreading and the emergence of the Mpemba effect for generic $q$ at $t\sim N_A$. 
\end{itemize}

\section{Additional details on $U(1)$ symmetric random circuits and their numerical implementation}
Throughout this work, we study qudits described by the local Hilbert space $\mathcal{H}_{2,q}\simeq \mathbb{C}^2\otimes \mathbb{C}^q$ of dimension $d=2q$, with $q=1$ corresponding to qubits. 
A basis of the local Hilbert space is given by $\{|x\rangle\equiv |\sigma,\alpha\rangle\}$, with $\sigma =0,1$ labeled as the spin degree of freedom, while $\alpha=0,\dots,q-1$ is a color (or flavor) degree of freedom. 
It is natural to pass to the doubled Hilbert space formulation. 
Any operator $A=\sum_{x,y} A_{x,y}|x\rangle\langle y |$ acting on the many-body system $\mathcal{H} = \mathcal{H}_{2,q}^{\otimes N}$, is expressible as a vector $|A\rrangle = \sum_{x,y} A_{x,y} |x,y\rrangle$. With this definition we have $\mathrm{tr}(B^\dagger A) \mapsto \llangle B | A\rrangle$, and $UAU^\dagger \mapsto (U\otimes U^*)|A\rrangle$. 

As discussed in the Main Text, the key ingredient to investigate $\mathcal{P}^{(2)}_A$ and $\mathcal{P}^{(2)}_{A,Q}$ is the Haar average of two body gates $\mathcal{T}_{i,i+1}=\mathbb{E}_\mathrm{Haar}\left[(U_{i,i+1} \otimes U_{i,i+1}^*)^{\otimes 2}\right]$. Using the decomposition in symmetry sectors, cf. Main Text: we have
\begin{equation}
\begin{split}
    \mathcal{T}_{i,i+1} &= \sum_{S_1 \neq S_2} \left[\mathbb{E}_{\mathrm{Haar}}\left({(u^{(S_1)}_{i,i+1})^\ast \otimes u^{(S_2)}_{i,i+1}  \otimes (u^{(S_2)}_{i,i+1})^\ast  \otimes u^{(S_1)}_{i,i+1} }\right)+
\mathbb{E}_{\mathrm{Haar}}\left({(u^{(S_1)}_{i,i+1})^\ast \otimes u^{(S_1)}_{i,i+1}  \otimes (u^{(S_2)}_{i,i+1})^\ast  \otimes u^{(S_2)}_{i,i+1} }\right)\right]+\\\qquad 
 &\sum_{S}\mathbb{E}_{\mathrm{Haar}}\left({(u^{(S)}_{i,i+1})^\ast \otimes u^{(S)}_{i,i+1}  \otimes (u^{(S)}_{i,i+1})^\ast  \otimes u^{(S)}_{i,i+1} }\right)\;,
\end{split}
\end{equation}
where the sum over $S,S_1,S_2 = -1,0,1$ is over the symmetry sectors of the local charge $Z_{i,i+1}$.
Let us define the two replica states on two sites
\begin{align}
& |{I_{S_1,S_2}^{+}}\rrangle = \sum_{ x_1, x'_1, x_2, x'_2} |{x_1 x_1 x_2 x_2}\rrangle |{x_1' x_1' x_2' x_2'}\rrangle \delta_{s_1+s_1',S_1}
 \delta_{s_2+s_2',S_2}\\
&  |{I_{S_1,S_2}^{-}}\rrangle = \sum_{ x_1, x'_1, x_2, x'_2} |{x_1 x_2 x_2 x_1}\rrangle|{x_1' x_2' x_2' x_1'}\rrangle \delta_{s_1+s_1',S_1}
 \delta_{s_2+s_2',S_2}\;,
\end{align}
where $s_i$ is the magnetization of the spin degree of freedom $\sigma_i$ in $x_i$, with $s_i=+1$ for $\sigma_i=0$ and $s_i=-1$ for $\sigma_i=1$. Performing the Haar average using the Weingarten calculus~\cite{Collins2006}, we have for $q>1$
\begin{align}
& \mathbb{E}_{\mathrm{Haar}}\left({U^\ast_{S_1} \otimes U^{\ }_{S_2} \otimes U^\ast_{S_2} \otimes U^{\ }_{S_1}}\right) = \frac{1}{d_{S_1} d_{S_2}} |{I_{S_1, S_2}^{-}}\rrangle \llangle{I_{S_1, S_2}^{-}}|  \\
 & \mathbb{E}_{\mathrm{Haar}}\left({U^\ast_{S_1} \otimes U^{\ }_{S_1} \otimes U^\ast_{S_2} \otimes U^{\ }_{S_2}}\right) = \frac{1}{d_{S_1} d_{S_2}} |{I_{S_1, S_2}^{+}}\rrangle \llangle{I_{S_1, S_2}^{+}}| \\
&\mathbb{E}_{\mathrm{Haar}}\left({U^\ast_{S} \otimes U^{\ }_{S} \otimes U^\ast_{S} \otimes U^{\ }_{S}}\right) = \frac{1}{d_{S}^2-1} \left[
|{I_{S, S}^{+}}\rrangle \llangle{I_{S, S}^{+}}|+
|{I_{S, S}^{-}}\rrangle \llangle{I_{S, S}^{-}}|-
\frac{1}{d_S q^2}
\left(|{I_{S, S}^{+}}\rrangle \llangle{I_{S, S}^{-}}|
+
|{I_{S, S}^{-}}\rrangle \llangle{I_{S, S}^{+}}|
\right)
\right]\;.
\end{align}
We can recast Eq.~\eqref{eq:transf} using the following replica states on each qudit
\begin{subequations}
\label{eq:murbdef}
\begin{align}
    |+,r,b\rrangle &= \sum_{x_1,x_2} |x_1,x_1,x_2,x_2\rrangle \delta_{s_1,r}\delta_{s_2,b}\\
    |-,r,b\rrangle &= \sum_{x_1,x_2} |x_1,x_2,x_2,x_1\rrangle \delta_{s_1,r}\delta_{s_2,b}
\end{align}
\end{subequations}
We note that $\llangle{\pm,r,b | \pm,r',b'} \rrangle = \delta_{r,r'} \delta_{b,b'} q^2 $ and $ \llangle{\pm, -\frac12,-\frac12  | \mp, -\frac12,-\frac12 }\rrangle= \llangle{\pm,  +\frac12,+\frac12   | \mp, +\frac12,+\frac12 }\rrangle = q$. 
With these definitions, we note that
\begin{equation}
 |{I^{\pm}_{S_1, S_2}}\rrangle = \sum_{\substack{r, b, r', b'}} |{\pm, r,b}\rrangle |{\pm,r',b'} \rrangle \delta_{r+r', S_1}\delta_{b+b', S_2}.
\end{equation}
Combining all these results, we obtain the expression $\mathcal{T}_{i,i+1}$ presented in the Main Text for $q\ge 2$. 

In the limit $q\to 1$, we notice the degeneracy $|\pm, -\frac12,-\frac12 \rrangle = |\mp, -\frac12,-\frac12 \rrangle$ and $|\pm, +\frac12,+\frac12 \rrangle = |\mp, +\frac12,+\frac12 \rrangle$, that reduces the local replica Hilbert space to a $6$ dimensional manifold, see also Ref.~\cite{keyserlingk2018}.

\subsection{Numerical implementation}
The implementation codes (and the numerical data) are available in Ref.~\cite{dataavail}. 
We have two independent implementations. On the one hand, we construct the exact dynamics on the full Hilbert space. This requires writing the states as arrays with $(2q)^{N}$ elements. Gates are $(2q)^2 \times (2q)^2$ matrices acting on the state through sparse matrix multiplication. 
The reduced density matrix $\rho_A = X X^\dagger$, where  $|\Psi\rangle \mapsto X$ is the reshaping of the state as a matrix of dimension $(2q)^{N_A}\times (2q)^{N-N_A}$. Similarly, the decohered density matrix $\rho_{A,Q}$ is obtained setting the symmetry-breaking elements in $\rho_A$ to zero.
The entropic measures are then obtained via exact diagonalization. 
We compute the quench and annealed average asymmetric collecting $\mathcal{N}_\mathrm{realizations}\ge 10^4$ realizations of the circuits, ranging up to $N\le 24$ for $q=1$ and $N\le 12$ for $q=2$. 

The tensor network implementation for the R\'enyi 2 entanglement asymmetry is built upon the replica trick and the Haar average, employing the library \textsc{ITensor}~\cite{itensor}. 
Given the effective Hilbert space dimension $d_\mathrm{eff} = 6$ for $q=1$ and $d_\mathrm{eff}=8$ for $q\ge 2$, we write the initial and final states entering $\mathcal{P}_A^{(2)}$ and $\mathcal{P}_{A,Q}^{(2)}$ as matrix-product states (MPS)~\cite{Schollw_ck_2011}. For notational convenience, we relabel the vectors $|\mu_i,r_i,b_i\rrangle$ as $|\omega\rrangle$ with $\omega_i=0,\dots,d_\mathrm{eff}-1$.

We note that $|\rho_0^{\otimes 2}\rrangle$ has elements outside the effective Hilbert space $\mathcal{H}_\mathrm{replica} = \mathrm{span}\{\omega_i\}$. Nevertheless, they will be projected out after a single layer of $\mathcal{T}_{i,i+1}$. (We have verified this fact implementing the tensor contraction $\mathcal{T}_{i,i+1}$ as a $(2q)^8\otimes (2q)^8$ projector onto $|\rho_0^{\otimes2}\rrangle$ for $q=1$ and $q=2$, albeit we expect this to hold in general). 
As a result, we can study the replica problem directly in the space $\mathcal{H}_\mathrm{replica}$ implementing the initial state $\mathfrak{P} |\rho_0^{\otimes 2}\rrangle$ with $\mathfrak{P}$ is the projection onto $\mathcal{H}_\mathrm{replica}$. 
We have
\begin{equation}
    \mathfrak{P}|\rho_0^{\otimes2}\rrangle = \sum_{\omega_1,\dots,\omega_N=0}^{d_\mathrm{eff}-1} A_1^{\omega_1} A_2^{\omega_2} \cdots A_N^{\omega_N} |\omega_1,\dots,\omega_N\rrangle\;,
\end{equation}
with, for any $\omega_k$, $A_k^{\omega_k}$ a $\chi\times\chi$ matrix for $k=2,3,\dots,N-1$, and $A_1^{\omega_1}$ ($A_N^{\omega_N}$) a $1\times \chi$ ($\chi\times 1$) matrix. Here $\chi$ is the bond dimension numbering the virtual indices. 
In a similar fashion, we represent the final states for the purity $\mathcal{P}_A^{(2)}$ and decohered purity $\mathcal{P}_{A,Q}^{(2)}$, respectively
\begin{equation}
\begin{split}
        |\mathcal{F}_A\rrangle &= \sum_{\omega_1,\dots,\omega_N=0}^{d_\mathrm{eff}-1} B_1^{\omega_1} B_2^{\omega_2} \cdots B_N^{\omega_N} |\omega_1,\dots,\omega_N\rrangle\;,\\
    |\mathcal{F}_{A,k}\rrangle&= \sum_{\omega_1,\dots,\omega_N=0}^{d_\mathrm{eff}-1} B^{\omega_1}_1(k) B^{\omega_2}_2(k)\cdots B^{\omega_N}_N(k) |\omega_1,\dots,\omega_N\rrangle\;\;.
\end{split}
\end{equation}
Analogously, we represent the matrix product operator $\mathcal{T}$, cf. Main Text, acting on the effective degrees of freedom $\omega_k$. These are all the ingredients to compute the annealed averaged entanglement asymmetry and reproduce the results presented in the Main Text.
For the numerical simulations we have fixed $\varepsilon = 10^{-15}$ as the truncation error for the singular value decomposition required by the compression algorithms~\cite{itensor,Schollw_ck_2011}, and let the bond dimension grow without bounds (we did not require to fix a maximal bond dimension for the timescales and system sizes explored in our work).

\section{Infinite $q$ limit}
In this section, we discuss the limit of $q\to\infty$ in the thermodynamic limit of the system. First, we shall discuss how this limit allow us to pinpoint a Markov process. Then, we discuss how the scaling limit allows for analytical insights at times $t\lesssim N_A$. 
\subsection{Markov process
\label{sec:markovianlargeq}}
For $N\to\infty$ the decoherence channel $\mathcal{C}_A(\bullet)$ in the Main Text can be written in the doubled space, as
\begin{equation}
    \mathcal{C} = \int_{-\pi}^{\pi} \frac{dk}{2\pi} e^{ik/2 \sum_{j\in A} Z_j}\otimes e^{-ik/2 \sum_{j\in A} Z_j} = \int_{-\pi}^{\pi} \frac{dk}{2\pi} \cos(k/2)^{2N_A} 
    \prod_{j\in A}\left[(1+i\tan(k/2)Z_{j,r}) (1-i\tan(k/2)Z_{j,b})\right]
\label{eq:zio}
\end{equation}
where we used that the charges on two eigenstates always differ by an integer, and we denote $Z_{j,\alpha}$ as Pauli $z$ operators acting on site $j$ and replica $\alpha$. 
For consistency, we refer to the ``ket'' degrees of freedom $\alpha=r$ as the red ones and the ``bra'' degrees of freedom $\alpha=b$ as the blue ones. 
It is instructive to study the initial time first. In this case $\rho_0$ is a product state, and therefore $\mathcal{P}_A^{(2)}(t=0)=\llangle \mathcal{F}_A |\rho_0^{\otimes 2}\rrangle =1$. Instead we have 
\begin{equation}
    \mathcal{P}_{A,Q}^{(2)}(t=0)=\llangle \mathcal{F}_A | \mathcal{C}_A \otimes \mathcal{C}_A |\rho_0^{\otimes 2}\rrangle = \llangle \rho_0 |  \mathcal{C}_A | \rho_0\rrangle = \int \frac{dk}{2\pi} e^{N_A f_0(k,\theta)},
\end{equation}
with $f_0(k,\theta) = \ln[\cos(k/2)^2(1+\tan^2(k/2)\cos^2(\theta))]$.
For large $N_A$, we evaluate the integral by steepest descent. Since the maximum value of $f_0(k,\theta)$ is at $k=0$, we have 
\begin{equation}
\label{eq:saddlet0}
   \mathcal{P}_{A,Q}^{(2)}(t=0) \simeq \int_{-\infty}^\infty \frac{dk}{2\pi } e^{- {N_A k^2\sin^2(\theta)}/{4}} = \frac{1}{\sqrt{\pi N_A} \sin\theta},
\end{equation}
recasting the result in~\cite{ares2022entanglement}.
At later time, the unitary evolution will generate non-trivial interactions between the replicae, and the statistical mechanics model will become more complicated.
In the following, we specialize to the case that $A=\{1,\dots,N_A\}$ and the complement $A_c$ is the rest. 
It is useful to introduce the dressed swap operators,  
\begin{equation}
    \mathcal{F}_{\rr,\bb}(x) = \prod_{j<N_A} \left(Z_j^{n_r(j)}\otimes Z_j^{n_b(j)}\right) \mathcal{F}(x),
\end{equation}
where $\rr, \bb$ indicate a certain binary configuration of occupation numbers on each site, $n_{\alpha}(j) \in \{0,1\}$ for $\alpha = r,b$.
With this notation, we have 
\begin{equation}
    \mathcal{P}_A^{(2)}(t) = \llangle \mathcal{F}_{\mathbf{0},\mathbf{0}}(N_A) |\mathcal{T}^t |\rho_0^{\otimes 2}\rrangle \;.
\end{equation}
We then introduce the operator
\begin{equation}
    G(z_r,z_b;x) = \sum_{\rr,\bb} z_r^{N_r} z_b^{N_b} \mathcal{F}_{\rr,\bb}(x),
\end{equation}
where $N_{r/b} = \sum_{j} n_{r/b}(j)$ are the total number of red/blue particles in each configuration, while $z_r$ and $z_b$ play the role of fugacities of the two particle species. 
Using the decomposition~\eqref{eq:zio}, we can express the overlap as 
\begin{equation}
\label{eq:troverlap}
    \mathcal{P}_{A,Q}^{(2)}= 
    \int_{-\pi}^{\pi} \frac{dk}{2\pi} \cos(k/2)^{2N_A} 
    \llangle G(i \tan(k/2),- i \tan(k/2);N_A)| \mathcal{T}^t |\rho_0^{\otimes 2}\rrangle\;.
\end{equation}
The convenience of this description comes when taking the $q\to\infty $ limit. The transfer matrix simplifies considerably, and the operators $\mathcal{F}_{\rr, \bb}(x)$ evolve into combinations of the others, giving rise to Markovian dynamics with a set of probabilistic rules. 
We refer to Ref.~\cite{rakovszky2019} for the details and report here the rules.

\begin{figure}
    \centering
    \includegraphics{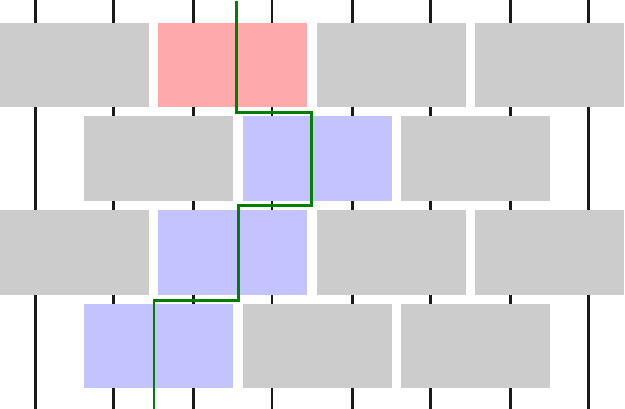}
    \caption{A schematic representation of the stochastic dynamics describing the $q\to \infty$ limit. The transition probability on each pair of sites is described by three $16 \times 16$ matrices, $M, L, R$ connecting two configurations $(n_r(j), n_b(j), n_r(j+1), n_b(j+1)) \to (n_r'(j), n_b'(j), n_r'(j+1), n_b'(j+1))$. The matrix $M$ (gray) is applied for a pair of sites away from the interface $x \neq j+1/2$; on the contrary, the pair of sites crossed by the random walker evolves with $L$ (blue) or $B$ (red) if the walker moves left or right
    \label{fig:circuitinterface}}
\end{figure}

We define a configuration by classical stochastic variables: the occupation number of "red" and "blue" particles $n_r(j)$ and $n_b(j)$, with $n_{r/b}(j) \in \{0,1\}$ and $j = 1,\ldots, N$, and the position of the interface $x$. As the interface is formally sitting on the bonds, we set its coordinate as $J = j + 1/2$, referring to the interface between $j$ and $j+1$. The full configuration $(n_r, n_b, x)$ is evolved according to a simple Markov process made out of two-site stochastic gates and a brick wall geometry (see Fig.~\ref{fig:circuitinterface}). There are three stochastic matrices $M, L ,R$ describing the transition probabilities from the two-site local configuration of the blue/red particles. 
Considering the pair at $j, j+1$, we have three cases: either $J \neq j + 1/2$ and in such case, the interface does not move, and the configuration of particles evolves according to the probabilities in the matrix $M_{j,j+1}$. Instead, if $J= j + 1/2$, then the interface moves one step left or right with equal probabilities, and the configuration of the particles is evolved according to the matrices $L_{j,j+1}$ and $R_{j,j+1}$ respectively.
More explicitly, denoting as $A = (n_r(j), n_b(j), n_r(j+1), n_b(j+1))$ and $A' = (n_r'(j), n_b'(j), n_r'(j+1), n_b'(j+1))$ the local configurations of red/blue particles of two neighboring sites, we have the transition probability
\begin{equation}
    \operatorname{Prob}([A,x] \to [A',x']) = 
    \begin{cases}
        [M_{j,j+1}]_{A',A} \delta_{x,x'}\;, & x \neq j + 1/2\\
        \frac{1}{2} [L_{j,j+1}]_{A',A} \delta_{x', x - 1} \;, & x = j + 1/2 \\
        \frac{1}{2} [R_{j,j+1}]_{A',A} \delta_{x', x + 1} \;, & x = j - 1/2\,.
    \end{cases}
\end{equation}
The explicit form of the matrices $M,L,R$ can be written compactly using Pauli operators, $\sigma_{j}^{\mu, \alpha}$, with $\alpha = r,b$ and $\mu = x,y,z$. 
We also introduce the operators
\begin{align}
    &\mathcal{S}_{j,j+1}^{(\alpha)}
    = \frac{1}{2}(1+\sigma _{j}^{x,\alpha} \sigma _{j+1}^{x,\alpha} + \sigma_{j}^{y,\alpha} \sigma _{j+1}^{y,\alpha} + \sigma _{j}^{z,\alpha} \sigma _{j+1}^{z,\alpha})  \;, \\
    &\mathcal{L}_{j,j+1}^{(\alpha)} = \frac{1}{4}\left[(3 \sigma^{x,\alpha}_j
    + \sigma^{x,\alpha}_{j+1}
    +i (\sigma^{y,\alpha}_j
    \sigma^{z,\alpha}_{j+1}-
    \sigma^{z,\alpha}_j
    \sigma^{y,\alpha}_{j+1}
    )\right] \;, \\
    &\mathcal{R}_{j,j+1}^{(\alpha)} = 
    \mathcal{S}_{j,j+1}^{(\alpha)}
    \mathcal{L}_{j,j+1}^{(\alpha)}
    \mathcal{S}_{j,j+1}^{(\alpha)},
\end{align}
where $\mathcal{S}_{j,j+1}^{(\alpha)}$ is just the swap operators exchanging the state of the particles of type $\alpha$ on sites $j,j+1$.
One then has
\begin{align}
    &M_{j,j+1}=
    \frac 1 4 (\mathbb{1}_{j,j+1} + \mathcal{S}_{j,j+1}^{(r)})
    (\mathbb{1}_{j,j+1} + \mathcal{S}_{j,j+1}^{(b)}),
    \label{eq:Mmatrix}
    \\
    &R_{j,j+1} = \frac{1}{2} (M_{j,j+1} + \mathcal{R}_{j,j+1}^{(r)}\mathcal{R}_{j,j+1}^{(b)}), \\
    &L_{j,j+1} = \frac{1}{2} (M_{j,j+1} + \mathcal{L}_{j,j+1}^{(r)}\mathcal{L}_{j,j+1}^{(b)}).
\end{align}
We note that the matrix $M$ describes a (discrete time) symmetric simple exclusion process (SSEP) independently for both red and blue particles. In other words, each particle can hop left/right with equal probabilities, with the only constraint that two particles of the same color cannot occupy the same site.  
The matrices $L/R$ describe more complex dynamics involving the creation/destruction of particles always in pairs. This fact leads to a non-trivial coarse-grained picture.

\subsection{Coarse-grained picture at $q \to \infty$}
Let us consider the coarse-grained picture that emerges using the macroscopic fluctuation theory (MFT). We assume the interval $N_A \gg t \gg 1$, so the boundaries are unimportant. Following \cite{Derrida2009}, we consider diffusive scaling of the coordinates on the relevant length scale $\sqrt{t}$,
\begin{equation}
\label{eq:coarsegrainedvars}
    x = j/\sqrt{t} \;, \qquad \tau = t'/t,
\end{equation}
with $x \in (-\infty, \infty)$ and $\tau \in [0,1]$. Then, we can introduce coarse-grained density as
\begin{equation}
    \rho_{\alpha}(x, \tau) = n_{\alpha}(x t^2, \tau t) \;, \qquad \alpha \in \{r,b\}.
\end{equation}
Let us first discuss the dynamics away from the interface. In this case, the stochastic evolution of red and blue particles is controlled by matrix $M$ in Eq.~\eqref{eq:Mmatrix}. As we explained, this corresponds to the simple process where, at each discrete time step, each particle hops left/right with equal probability as long as the destination does not contain a particle of the same type. In the coarse-grained scaling~\eqref{eq:coarsegrainedvars}, one has the description in terms of two independent SSEP described in MFT by the probability of a given density and current profile
\begin{equation}
\label{eq:MFT}
    \operatorname{Pro}(\rho(x,\tau), j(x,\tau)) = \exp\left[-\sqrt{t} \int_0^1 d\tau \int_{-\infty}^\infty \frac{(j(x,\tau) + D\partial_x \rho)^2}{2 \sigma(\rho)} \right] \delta(\partial_\tau \rho + \partial_x j),
\end{equation}
where the diffusion constant and the mobility take the form
\begin{equation}
\label{eq:DsigmaSSEP}
    D(\rho) = 1 \;, \qquad \sigma(\rho) = 2\rho(1-\rho) \;.
\end{equation}

Let us now consider the effect of the interface. First, its dynamics in discrete time is a simple random walk, so $J(t) = \delta_1 + \delta_2 + \ldots + \delta_{2t}$, where $\delta_i \in \{-1,1\}$ are the increments at each step. Once the scaling \eqref{eq:coarsegrainedvars} is used, we denote the rescaled coordinate of the interface as
 \begin{equation}
x(\tau) = t^{-1/2} J(\tau t) 
 \end{equation}
with this choice, because of the central limit theorem, the variable $x(\tau)$ converges at large $t \to \infty$ to a Gaussian random variable with zero average and variance $2\tau$. Since its increments are uncorrelated, $x(\tau)$ converges to a Wiener process.
Red and blue particles can be created and destroyed in proximity to the interface $x \sim x(\tau)$. The two processes occur at the same rate, so we assume that, in coarse-grained scaling, the effect of the membrane is to inject/remove particles so that the density 
\begin{equation}
\label{eq:MFTbound}
    \rho_\alpha(x = x(\tau), \tau) \sim 1/2.
\end{equation}
The approximate symbol here is used to clarify that the density is still allowed to fluctuate microscopically~\cite{Saha_2023}.

\subsection{Low-density expansion}

To simplify our treatment, we will ignore the fluctuations of the variable $x(\tau)$ and assume we can set $x(\tau) = 0$, so that we have the boundary condition in space
\begin{equation}
\label{eq:walldens}
    \rho_\alpha(0, \tau) \sim \rhoa \;,
\end{equation}
where in our case $\rhoa = 1/2$. Additionally, we are interested (see below) in the case where a fixed density of particles is present at the initial time inside the interval, so 
\begin{equation}
\label{eq:inidens}
    \rho_\alpha(x, 0) \sim \begin{cases}
        \rhob & x>0 \;, \\
        0 & x<0 \;.
    \end{cases} 
\end{equation}
The $\sim$ in Eq.~\eqref{eq:inidens} refers to the fact that the initial density is kept statistically fixed but still allows for thermodynamic fluctuations. This is known as \textit{annealed} average in this context~\cite{Derrida2009}.
Because of Eq.~\eqref{eq:walldens},
red and blue particles decouple and the dynamics for $x>0$ and $x<0$.  Since each particle (irrespectively of its color) present at final time contributes with a factor $\cos(\theta)$ when evaluated on the initial state, we are interested in computing
\begin{equation}
\label{eq:purityfunctional}
    \mu(\lambda; \rhoa, \rhob) := \lim_{t\to\infty} \frac{1}{\sqrt{t}} \ln \int \mathcal{D}\rho \mathcal{D}j 
    \;    \operatorname{Pro}(\rho(x,\tau), j(x,\tau)) e^{\lambda \sqrt{t} \int_{0}^\infty dx (\rho(x, 1) - \rho(x,0))} \;,
\end{equation}
where the functional integral are rescricted to $x>0$ with the boundary conditions (\ref{eq:walldens}, \ref{eq:inidens}). 

\begin{figure*}[t]
    \centering
    \includegraphics[width=0.7\textwidth]{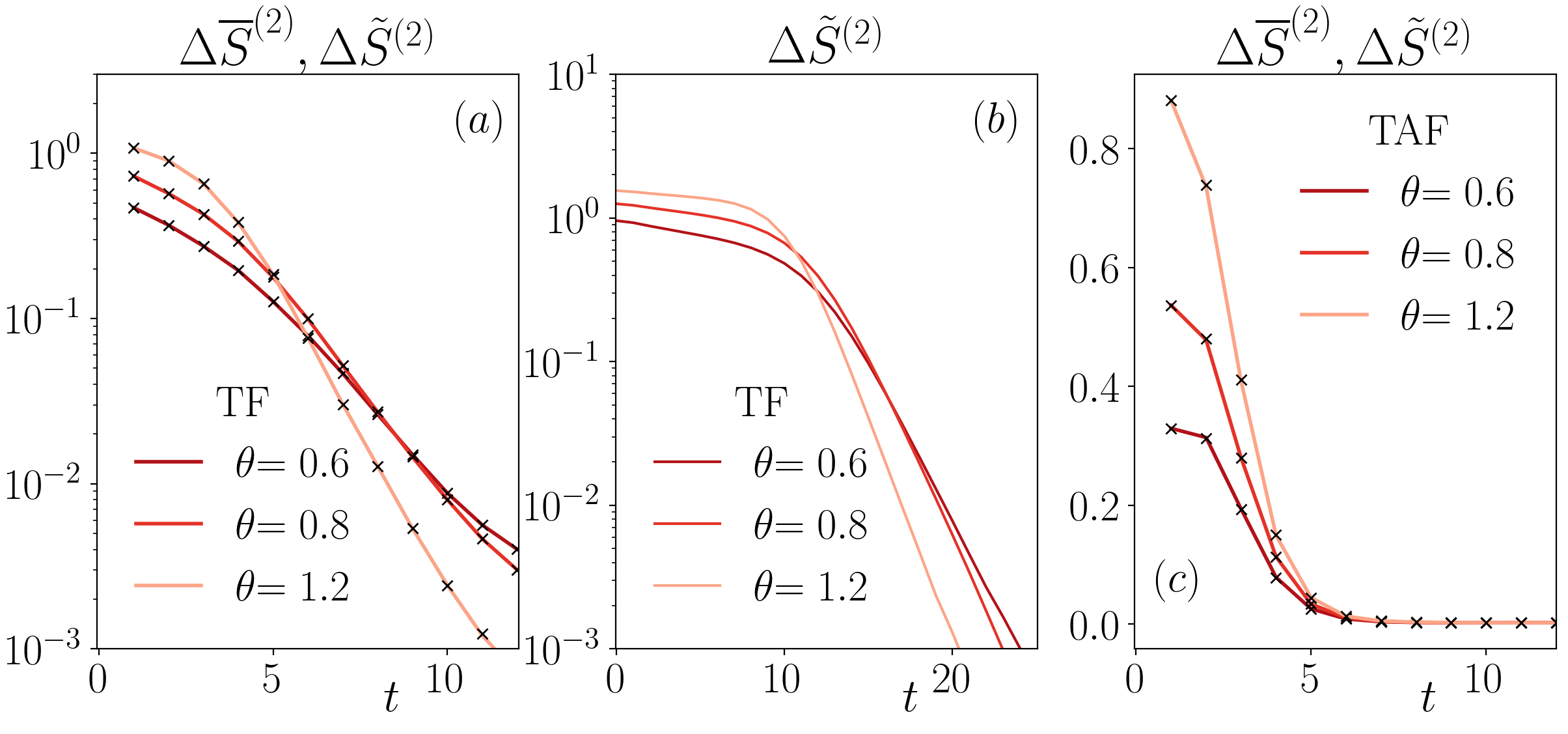}
    \caption{Numerical results for $q=2$. (a) Quench (solid lines) and annealed averaged (crosses)  R\'enyi 2 entanglement asymmetry starting from the TF state, with $N=12$ and $N_A=3$, and averaging over $\mathcal{N}=10^3$ circuits. The data shows that there is QME even at $q=2$.  
    (b) Annealed averaged R\'enyi 2 entanglement asymmetry starting from the TF state for $N=128$ and $N_A=8$, as obtained with tensor networks. 
    (c) Starting from the TAF state, the Mpemba effect is absent, as for $q=1$. The data compares the quench (solid lines) and the annealed (crosses) values of $\Delta S^{(2)}$ obtained with exact methods for $N=8$ and $N_A=2$. }
    \label{fig:mpemba2}
\end{figure*}

In our case we will have to set eventually $\lambda = \ln \cos \theta$ corresponding to evaluating the final density of red/blue particles in the initial state. 
This problem was recently considered in~\cite{Saha_2023}, where different scaling of the rates imposing Eq.~\eqref{eq:walldens} were considered. Our case corresponds to the fast-coupling limit. Nonetheless, the exact form of $\mu(\lambda; \rhoa, \rhob)$ for arbitrary $\lambda$ is not known. At linear order in $\rhoa, \rhob$ one has $\mu(\lambda; \rhoa, \rhob) = \mu_0(\lambda, \rhoa, \rhob) +  O(\rhoa^2, \rhob^2, \rhoa \rhob)$, with the simple form
\begin{equation}
\mu_0(\lambda; \rhoa, \rhob) = \frac{2 (\rhoa (e^\lambda-1)+\rhob(e^{-\lambda}-1))}{\sqrt{\pi }},
\end{equation}
corresponding to replacing SSEP with non-interacting Brownian motions. However, we are interested in $\rhob = 1/2$, away from this non-interacting regime. While a full form is unavailable, a few orders in powers of $\lambda$ are known. Setting
\begin{equation}
\mu(\lambda; \rhoa, \rhob) =
\mu_0(\lambda; \rhoa, \rhob) + \sum_{k=1}^\infty \frac{\lambda^k c_k(\rhoa, \rhob)}{k!} ,
\end{equation}
where the first coefficients take the form
\begin{align}
   c_1(\rhoa, \rhob) &= 0\;, \qquad
   c_2(\rhoa, \rhob) = 
   -\frac{4}{\sqrt{\pi}} [(\sqrt{2}-1)(\rhoa^2 + \rhob^2) + (3 - 2 \sqrt{2})\rhoa \rhob],
  \\
   c_3(\rhoa, \rhob) &= \frac{4 (\rhoa-\rhob)}{3\sqrt{\pi}} [9 (1-\sqrt{2}) (\rhoa+\rhob-2 \rhoa \rhob)-2 (9 \sqrt{2}-8 \sqrt{3}) (\rhoa-\rhob)^2].
\end{align}
We can use directly this result to derive the behavior of $\mathcal{P}_A^{(2)}$ and $\mathcal{P}_{A,Q}^{(2)}$ perturbatively at small $\theta$.
For the purity, we simply have
\begin{equation}
        \lim_{t \to \infty} \frac{1}{\sqrt{t}}\ln \mathrm{tr}[\rho_{A,t}^2] \simeq 4 \mu(\lambda = \ln\cos\theta; 1/2,0) =
        -\frac{2 \theta ^2}{\sqrt{\pi }}  
        +\frac{\left(4-3 \sqrt{2}\right) \theta ^4}{6 \sqrt{\pi }}+
        \frac{\left(15 \sqrt{2}-10 \sqrt{3}-4\right) \theta ^6}{45 \sqrt{\pi }} + O(\theta^8),
\end{equation}
where the limit of large $t$ is taken still assuming $t \ll N_A$. The factor $4$ comes from the contribution of red/blue particles and $x>0$, $x<0$.

The calculation of $\mathcal{P}_{A,Q}^{(2)}$ is slightly more involved. We note that the fugacities $z_\alpha$ are practically enforcing initial densities $\rho_\alpha = z_\alpha/(1+z_\alpha)$ inside the interval $A$, leading  to
\begin{equation}
\label{eq:Gcoarse}
    \frac{\llangle G(z_r, z_b; x)| \mathcal{T}^t |\rho_0^{\otimes 2}\rrangle}{\llangle G(z_r, z_b; x)| \rho_0^{\otimes 2}\rrangle} \simeq \exp\left[\sqrt{t}( 2\mu(\lambda; 1/2, 0) + \mu\left(\lambda; 1/2, \frac{z_r}{1 + z_r}\right) + \mu\left(\lambda; 1/2, \frac{z_b}{1 + z_b}\right)\right]\;,
\end{equation}
in the coarse-grained picture.
Although the simplified Markovian dynamics presented in \ref{sec:markovianlargeq} is only valid when $t \ll N_A$, it is useful to set $N_A = \kappa \sqrt{t}$ and treat $\kappa$ as a generic parameter at large $t$. Plugging Eq.~\eqref{eq:Gcoarse} in \eqref{eq:troverlap}, with $z_{r/b} = \pm i \tan(k/2)$, we can evaluate the integral over $k$ at large $t$ via saddle point similarly to what was done in Eq.~\eqref{eq:saddlet0}. As the saddle point always lies at $k = 0$ for any $t$, we can expand 
\begin{equation}
        \frac{1}{\sqrt{t}}\ln 
        \llangle G(i \tan(k/2), - i \tan(k/2); x) | \mathcal{T}^t |\rho_0^{\otimes 2}\rrangle  = 4 \mu(\lambda; 1/2;0) + k^2 g(\theta)  ,
\end{equation}
where we set
\begin{equation}
    g(\theta) = \frac{\theta ^2}{2 \sqrt{\pi }}+\frac{\left(6 \sqrt{2}-5\right) \theta ^4}{12 \sqrt{\pi }}+\frac{\left(-22-165 \sqrt{2}+180 \sqrt{3}\right) \theta ^6}{360 \sqrt{\pi }}+O\left(\theta ^7\right).
\end{equation}
Finally, we can compute the integral over $k$ as
\begin{equation}
   \mathcal{P}_{A,Q}^{(2)} \simeq \mathcal{P}_{A}^{(2)} \int_{-\infty}^{\infty} \frac{dk}{2\pi} e^{- \sqrt{t} k^2  \left(\frac{\kappa} {4} \sin(\theta)^2  - g(\theta)\right)} = \frac{\mathcal{P}_{A}^{(2)} }{\sqrt{\pi (N_A \sin(\theta)^2- 4 \sqrt{t} g(\theta))}}.
\end{equation}
In conclusion, we get the annealed average entanglement asymmetry as 
\begin{equation}
    \Delta\tilde{S}_A^{(2)}(t)\simeq  1 - \frac{1}{\sqrt{\pi (N_A \sin(\theta)^2- 4 \sqrt{t} g(\theta))}} \simeq \Delta\tilde{S}_A^{(2)}(0) -\frac{2 \sqrt{t} g(\theta ) }{\sqrt{\pi N_A^3} \sin(\theta)^3}.
\end{equation}
This implies that the initial slope at which the asymmetry decays is $g(\theta) / \sin(\theta)^3$, which monotonously decreases with $\theta$. This result shows that the Mpemba effect is not visible in the short-time regime $t \ll N_A$.

\section{Additional numerical details}
We conclude the Supplemental material presenting numerical data for $q=2$, with local Hilbert space of dimension $d=4$. In Fig.~\ref{fig:mpemba2}(a) we present the quench and annealed average findings of $\Delta S_A^{(2)}$  for  $N=12$ and $N_A=3$ starting from the TF state $|\Psi_0^f(\theta)\rangle$. Despite the limited system sizes, the crossings demonstrate the emergence of QME.
Comparing quench (solid lines) and annealed average (crosses), Fig.~\ref{fig:mpemba2}(a) also highlights the self-averaging of the circuit even at higher $q$, a property that we believe to hold for any value of $q$. 
We further corroborate the exact computations by studying $N=128$ and $N_A=8$ via tensor networks, cf. Fig.~\ref{fig:mpemba2}(b). 
In Fig.~\ref{fig:mpemba2}(c), we show that, similarly to the $q=1$ case, even for $q=2$, the Mpemba phenomenology is absent when starting from the TAF initial state. 

\begin{figure*}[t]
    \centering
    \includegraphics[width=0.4\textwidth]{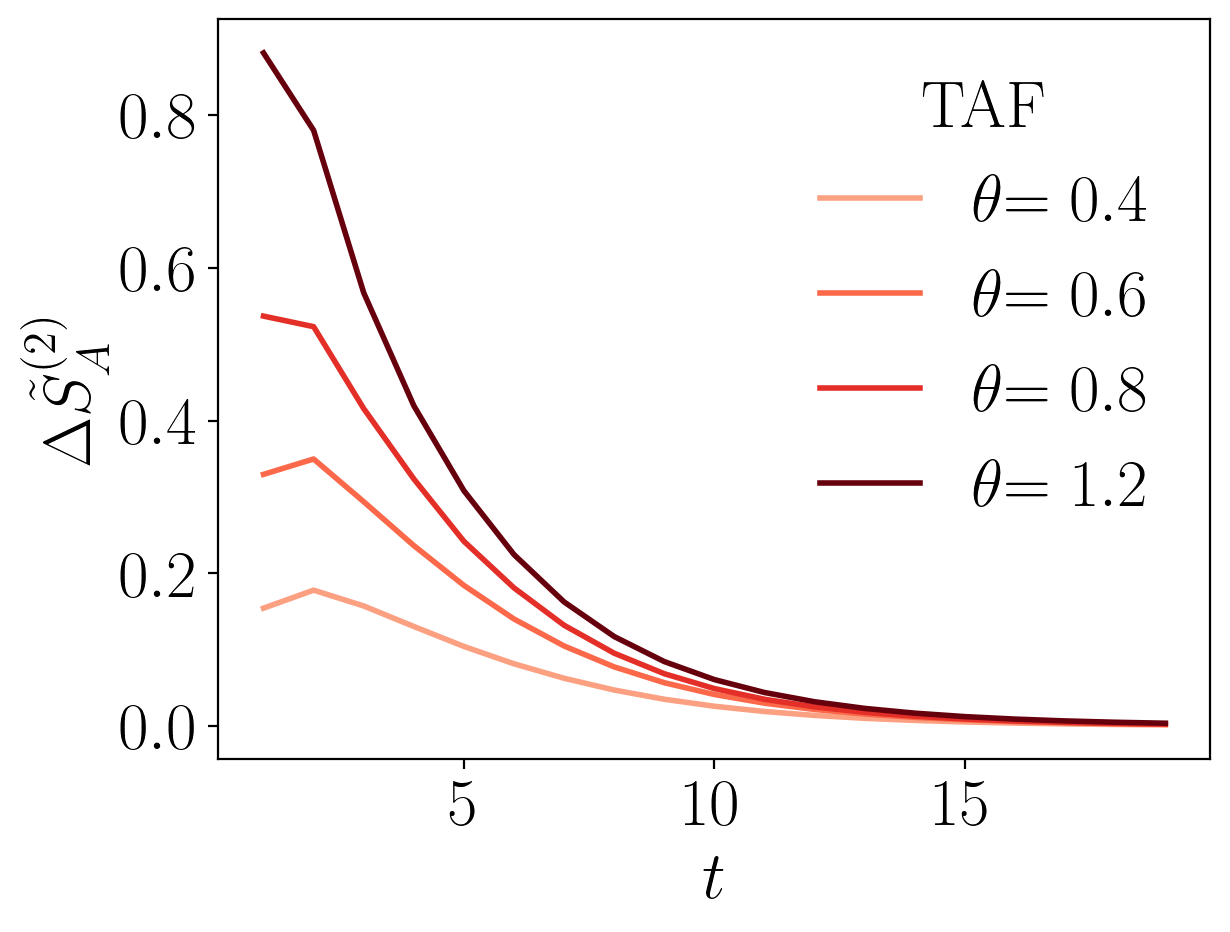}
    \caption{Absence of Mpemba effect for the TAF initial state for several $\theta$, $N_A=2$ and $N=256$.}
    \label{fig:ref}
\end{figure*}

In Fig.~\ref{fig:ref}, we finally demonstrate that the TAF state does not present Mpemba phenomenology even at large values of system size $N=256$ for $q=1$.

\end{document}

%% file: sketchmembrane.tex
\definecolor{mycolor}{rgb}{0.1,0.3,0.1}


    \draw[thick] (-4,0) rectangle (4,-5.2);
    \node at (-2, 0) [above] {$L_A$};
    \draw[line width=1.5] (-4,0) -- (0.5,0);

    \def\step{0.08}  
    \def\xstep{0.1}
    \def\branch_steps{64} 
    \def\steps{64}
    \coordinate (start) at (0,0); 

    \newcommand{\drawBrownianPath}[4]{
        \coordinate (current) at #1;
        \foreach \j in {1,...,#2} {
            \pgfmathsetmacro{\randstep}{rand} 
            \coordinate (next) at ($(current) + (2*\xstep*\randstep,-\step)$);
            \draw[thick, color=#4, opacity=#3] (current) -- (next);
            \coordinate (current) at (next); 
        }
    }

    \foreach \i in {1,...,\steps} {
        \pgfmathsetmacro{\rand}{rand}
        \coordinate (next) at ($(start) + (\rand*\xstep,-\step)$);
        \draw[line width=1.5, mycolor] (start) -- (next);
        \coordinate (start) at (next);
        
        \ifnum\i>0
         \pgfmathtruncatemacro{\imod}{mod(\i,8)}
                \ifnum\imod=1
                    \pgfmathsetmacro\opacity{0.2 + \i/\steps/2} 
                    \pgfmathsetmacro\stepstodo{64-\i}
                    \drawBrownianPath{(start)}{\stepstodo}{\opacity}{red};
                    \drawBrownianPath{(start)}{\stepstodo}{\opacity}{blue}
                \fi
        \fi
    }

    \draw[line width=1.5, mycolor, dashed] (0,0) -- (-4,-3);
        (0,0) rectangle (-4,-3);

%% file: remain2.bbl
\begin{thebibliography}{105}%
\makeatletter
\providecommand \@ifxundefined [1]{%
 \@ifx{#1\undefined}
}%
\providecommand \@ifnum [1]{%
 \ifnum #1\expandafter \@firstoftwo
 \else \expandafter \@secondoftwo
 \fi
}%
\providecommand \@ifx [1]{%
 \ifx #1\expandafter \@firstoftwo
 \else \expandafter \@secondoftwo
 \fi
}%
\providecommand \natexlab [1]{#1}%
\providecommand \enquote  [1]{``#1''}%
\providecommand \bibnamefont  [1]{#1}%
\providecommand \bibfnamefont [1]{#1}%
\providecommand \citenamefont [1]{#1}%
\providecommand \href@noop [0]{\@secondoftwo}%
\providecommand \href [0]{\begingroup \@sanitize@url \@href}%
\providecommand \@href[1]{\@@startlink{#1}\@@href}%
\providecommand \@@href[1]{\endgroup#1\@@endlink}%
\providecommand \@sanitize@url [0]{\catcode `\\12\catcode `\$12\catcode
  `\&12\catcode `\#12\catcode `\^12\catcode `\_12\catcode `\%12\relax}%
\providecommand \@@startlink[1]{}%
\providecommand \@@endlink[0]{}%
\providecommand \url  [0]{\begingroup\@sanitize@url \@url }%
\providecommand \@url [1]{\endgroup\@href {#1}{\urlprefix }}%
\providecommand \urlprefix  [0]{URL }%
\providecommand \Eprint [0]{\href }%
\providecommand \doibase [0]{https://doi.org/}%
\providecommand \selectlanguage [0]{\@gobble}%
\providecommand \bibinfo  [0]{\@secondoftwo}%
\providecommand \bibfield  [0]{\@secondoftwo}%
\providecommand \translation [1]{[#1]}%
\providecommand \BibitemOpen [0]{}%
\providecommand \bibitemStop [0]{}%
\providecommand \bibitemNoStop [0]{.\EOS\space}%
\providecommand \EOS [0]{\spacefactor3000\relax}%
\providecommand \BibitemShut  [1]{\csname bibitem#1\endcsname}%
\let\auto@bib@innerbib\@empty
\bibitem [{\citenamefont {Rigol}\ \emph {et~al.}(2008)\citenamefont {Rigol},
  \citenamefont {Dunjko},\ and\ \citenamefont {Olshanii}}]{rigol2008}%
  \BibitemOpen
  \bibfield  {author} {\bibinfo {author} {\bibfnamefont {M.}~\bibnamefont
  {Rigol}}, \bibinfo {author} {\bibfnamefont {V.}~\bibnamefont {Dunjko}},\ and\
  \bibinfo {author} {\bibfnamefont {M.}~\bibnamefont {Olshanii}},\ }\href
  {https://doi.org/10.1038/nature06838} {\bibfield  {journal} {\bibinfo
  {journal} {Nature}\ }\textbf {\bibinfo {volume} {452}},\ \bibinfo {pages}
  {854} (\bibinfo {year} {2008})}\BibitemShut {NoStop}%
\bibitem [{\citenamefont {Polkovnikov}\ \emph {et~al.}(2011)\citenamefont
  {Polkovnikov}, \citenamefont {Sengupta}, \citenamefont {Silva},\ and\
  \citenamefont {Vengalattore}}]{polkovnikov2011}%
  \BibitemOpen
  \bibfield  {author} {\bibinfo {author} {\bibfnamefont {A.}~\bibnamefont
  {Polkovnikov}}, \bibinfo {author} {\bibfnamefont {K.}~\bibnamefont
  {Sengupta}}, \bibinfo {author} {\bibfnamefont {A.}~\bibnamefont {Silva}},\
  and\ \bibinfo {author} {\bibfnamefont {M.}~\bibnamefont {Vengalattore}},\
  }\href {https://doi.org/10.1103/RevModPhys.83.863} {\bibfield  {journal}
  {\bibinfo  {journal} {Rev. Mod. Phys.}\ }\textbf {\bibinfo {volume} {83}},\
  \bibinfo {pages} {863} (\bibinfo {year} {2011})}\BibitemShut {NoStop}%
\bibitem [{\citenamefont {Fagotti}(2014)}]{fagotti2014}%
  \BibitemOpen
  \bibfield  {author} {\bibinfo {author} {\bibfnamefont {M.}~\bibnamefont
  {Fagotti}},\ }\href {https://dx.doi.org/10.1088/1742-5468/2014/03/P03016}
  {\bibfield  {journal} {\bibinfo  {journal} {J. Stat. Mech.}\ }\textbf
  {\bibinfo {volume} {2014}},\ \bibinfo {pages} {P03016} (\bibinfo {year}
  {2014})}\BibitemShut {NoStop}%
\bibitem [{\citenamefont {D’Alessio}\ \emph {et~al.}(2016)\citenamefont
  {D’Alessio}, \citenamefont {Kafri}, \citenamefont {Polkovnikov},\ and\
  \citenamefont {Rigol}}]{dalessio2016}%
  \BibitemOpen
  \bibfield  {author} {\bibinfo {author} {\bibfnamefont {L.}~\bibnamefont
  {D’Alessio}}, \bibinfo {author} {\bibfnamefont {Y.}~\bibnamefont {Kafri}},
  \bibinfo {author} {\bibfnamefont {A.}~\bibnamefont {Polkovnikov}},\ and\
  \bibinfo {author} {\bibfnamefont {M.}~\bibnamefont {Rigol}},\ }\href
  {http://dx.doi.org/10.1080/00018732.2016.1198134} {\bibfield  {journal}
  {\bibinfo  {journal} {Adv. Phys.}\ }\textbf {\bibinfo {volume} {65}},\
  \bibinfo {pages} {239} (\bibinfo {year} {2016})}\BibitemShut {NoStop}%
\bibitem [{\citenamefont {Calabrese}\ \emph {et~al.}(2016)\citenamefont
  {Calabrese}, \citenamefont {Essler},\ and\ \citenamefont
  {Mussardo}}]{calabrese2016}%
  \BibitemOpen
  \bibfield  {author} {\bibinfo {author} {\bibfnamefont {P.}~\bibnamefont
  {Calabrese}}, \bibinfo {author} {\bibfnamefont {F.~H.~L.}\ \bibnamefont
  {Essler}},\ and\ \bibinfo {author} {\bibfnamefont {G.}~\bibnamefont
  {Mussardo}},\ }\href {https://doi.org/10.1088/1742-5468/2016/06/064001}
  {\bibfield  {journal} {\bibinfo  {journal} {J. Stat. Mech.}\ }\textbf
  {\bibinfo {volume} {2016}},\ \bibinfo {pages} {064001} (\bibinfo {year}
  {2016})}\BibitemShut {NoStop}%
\bibitem [{\citenamefont {Essler}\ and\ \citenamefont
  {Fagotti}(2016)}]{essler2016}%
  \BibitemOpen
  \bibfield  {author} {\bibinfo {author} {\bibfnamefont {F.~H.~L.}\
  \bibnamefont {Essler}}\ and\ \bibinfo {author} {\bibfnamefont
  {M.}~\bibnamefont {Fagotti}},\ }\href
  {https://doi.org/10.1088/1742-5468/2016/06/064002} {\bibfield  {journal}
  {\bibinfo  {journal} {J. Stat. Mech.}\ }\textbf {\bibinfo {volume} {2016}},\
  \bibinfo {pages} {064002} (\bibinfo {year} {2016})}\BibitemShut {NoStop}%
\bibitem [{\citenamefont {Deutsch}(1991)}]{deutsch1991}%
  \BibitemOpen
  \bibfield  {author} {\bibinfo {author} {\bibfnamefont {J.~M.}\ \bibnamefont
  {Deutsch}},\ }\href {https://doi.org/10.1103/PhysRevA.43.2046} {\bibfield
  {journal} {\bibinfo  {journal} {Phys. Rev. A}\ }\textbf {\bibinfo {volume}
  {43}},\ \bibinfo {pages} {2046} (\bibinfo {year} {1991})}\BibitemShut
  {NoStop}%
\bibitem [{\citenamefont {Srednicki}(1994)}]{srednicki1994}%
  \BibitemOpen
  \bibfield  {author} {\bibinfo {author} {\bibfnamefont {M.}~\bibnamefont
  {Srednicki}},\ }\href {https://doi.org/10.1103/PhysRevE.50.888} {\bibfield
  {journal} {\bibinfo  {journal} {Phys. Rev. E}\ }\textbf {\bibinfo {volume}
  {50}},\ \bibinfo {pages} {888} (\bibinfo {year} {1994})}\BibitemShut
  {NoStop}%
\bibitem [{\citenamefont {Brenes}\ \emph {et~al.}(2021)\citenamefont {Brenes},
  \citenamefont {Pappalardi}, \citenamefont {Mitchison}, \citenamefont
  {Goold},\ and\ \citenamefont {Silva}}]{brenes2021}%
  \BibitemOpen
  \bibfield  {author} {\bibinfo {author} {\bibfnamefont {M.}~\bibnamefont
  {Brenes}}, \bibinfo {author} {\bibfnamefont {S.}~\bibnamefont {Pappalardi}},
  \bibinfo {author} {\bibfnamefont {M.~T.}\ \bibnamefont {Mitchison}}, \bibinfo
  {author} {\bibfnamefont {J.}~\bibnamefont {Goold}},\ and\ \bibinfo {author}
  {\bibfnamefont {A.}~\bibnamefont {Silva}},\ }\href
  {https://doi.org/10.1103/PhysRevE.104.034120} {\bibfield  {journal} {\bibinfo
   {journal} {Phys. Rev. E}\ }\textbf {\bibinfo {volume} {104}},\ \bibinfo
  {pages} {034120} (\bibinfo {year} {2021})}\BibitemShut {NoStop}%
\bibitem [{\citenamefont {Pappalardi}\ \emph {et~al.}(2022)\citenamefont
  {Pappalardi}, \citenamefont {Foini},\ and\ \citenamefont
  {Kurchan}}]{pappalardi2022}%
  \BibitemOpen
  \bibfield  {author} {\bibinfo {author} {\bibfnamefont {S.}~\bibnamefont
  {Pappalardi}}, \bibinfo {author} {\bibfnamefont {L.}~\bibnamefont {Foini}},\
  and\ \bibinfo {author} {\bibfnamefont {J.}~\bibnamefont {Kurchan}},\ }\href
  {https://doi.org/10.1103/PhysRevLett.129.170603} {\bibfield  {journal}
  {\bibinfo  {journal} {Phys. Rev. Lett.}\ }\textbf {\bibinfo {volume} {129}},\
  \bibinfo {pages} {170603} (\bibinfo {year} {2022})}\BibitemShut {NoStop}%
\bibitem [{\citenamefont {Pappalardi}\ \emph {et~al.}()\citenamefont
  {Pappalardi}, \citenamefont {Fritzsch},\ and\ \citenamefont
  {Prosen}}]{pappalardi2023}%
  \BibitemOpen
  \bibfield  {author} {\bibinfo {author} {\bibfnamefont {S.}~\bibnamefont
  {Pappalardi}}, \bibinfo {author} {\bibfnamefont {F.}~\bibnamefont
  {Fritzsch}},\ and\ \bibinfo {author} {\bibfnamefont {T.}~\bibnamefont
  {Prosen}},\ }\href@noop {} {}\Eprint {https://arxiv.org/abs/2303.00713}
  {arXiv:2303.00713} \BibitemShut {NoStop}%
\bibitem [{\citenamefont {Mpemba}\ and\ \citenamefont
  {Osborne}(1969)}]{Mpemba}%
  \BibitemOpen
  \bibfield  {author} {\bibinfo {author} {\bibfnamefont {E.~B.}\ \bibnamefont
  {Mpemba}}\ and\ \bibinfo {author} {\bibfnamefont {D.~G.}\ \bibnamefont
  {Osborne}},\ }\href {https://doi.org/10.1088/0031-9120/4/3/312} {\bibfield
  {journal} {\bibinfo  {journal} {Phys. Educ.}\ }\textbf {\bibinfo {volume}
  {4}},\ \bibinfo {pages} {172} (\bibinfo {year} {1969})}\BibitemShut {NoStop}%
\bibitem [{\citenamefont {Lasanta}\ \emph {et~al.}(2017)\citenamefont
  {Lasanta}, \citenamefont {Vega~Reyes}, \citenamefont {Prados},\ and\
  \citenamefont {Santos}}]{lasanta2017}%
  \BibitemOpen
  \bibfield  {author} {\bibinfo {author} {\bibfnamefont {A.}~\bibnamefont
  {Lasanta}}, \bibinfo {author} {\bibfnamefont {F.}~\bibnamefont {Vega~Reyes}},
  \bibinfo {author} {\bibfnamefont {A.}~\bibnamefont {Prados}},\ and\ \bibinfo
  {author} {\bibfnamefont {A.}~\bibnamefont {Santos}},\ }\href
  {https://doi.org/10.1103/PhysRevLett.119.148001} {\bibfield  {journal}
  {\bibinfo  {journal} {Phys. Rev. Lett.}\ }\textbf {\bibinfo {volume} {119}},\
  \bibinfo {pages} {148001} (\bibinfo {year} {2017})}\BibitemShut {NoStop}%
\bibitem [{\citenamefont {Lu}\ and\ \citenamefont {Raz}(2017)}]{lu2017}%
  \BibitemOpen
  \bibfield  {author} {\bibinfo {author} {\bibfnamefont {Z.}~\bibnamefont
  {Lu}}\ and\ \bibinfo {author} {\bibfnamefont {O.}~\bibnamefont {Raz}},\
  }\href {https://doi.org/10.1073/pnas.1701264114} {\bibfield  {journal}
  {\bibinfo  {journal} {PNAS}\ }\textbf {\bibinfo {volume} {114}},\ \bibinfo
  {pages} {5083} (\bibinfo {year} {2017})}\BibitemShut {NoStop}%
\bibitem [{\citenamefont {Klich}\ \emph {et~al.}(2019)\citenamefont {Klich},
  \citenamefont {Raz}, \citenamefont {Hirschberg},\ and\ \citenamefont
  {Vucelja}}]{klich2019}%
  \BibitemOpen
  \bibfield  {author} {\bibinfo {author} {\bibfnamefont {I.}~\bibnamefont
  {Klich}}, \bibinfo {author} {\bibfnamefont {O.}~\bibnamefont {Raz}}, \bibinfo
  {author} {\bibfnamefont {O.}~\bibnamefont {Hirschberg}},\ and\ \bibinfo
  {author} {\bibfnamefont {M.}~\bibnamefont {Vucelja}},\ }\href
  {https://doi.org/10.1103/PhysRevX.9.021060} {\bibfield  {journal} {\bibinfo
  {journal} {Phys. Rev. X}\ }\textbf {\bibinfo {volume} {9}},\ \bibinfo {pages}
  {021060} (\bibinfo {year} {2019})}\BibitemShut {NoStop}%
\bibitem [{\citenamefont {Kumar}\ and\ \citenamefont
  {Bechhoefer}(2020)}]{kumar2020}%
  \BibitemOpen
  \bibfield  {author} {\bibinfo {author} {\bibfnamefont {A.}~\bibnamefont
  {Kumar}}\ and\ \bibinfo {author} {\bibfnamefont {J.}~\bibnamefont
  {Bechhoefer}},\ }\href {https://doi.org/10.1038/s41586-020-2560-x} {\bibfield
   {journal} {\bibinfo  {journal} {Nature}\ }\textbf {\bibinfo {volume}
  {584}},\ \bibinfo {pages} {64} (\bibinfo {year} {2020})}\BibitemShut
  {NoStop}%
\bibitem [{\citenamefont {Bechhoefer}\ \emph {et~al.}(2021)\citenamefont
  {Bechhoefer}, \citenamefont {Kumar},\ and\ \citenamefont
  {Chétrite}}]{bechhoefer2021}%
  \BibitemOpen
  \bibfield  {author} {\bibinfo {author} {\bibfnamefont {J.}~\bibnamefont
  {Bechhoefer}}, \bibinfo {author} {\bibfnamefont {A.}~\bibnamefont {Kumar}},\
  and\ \bibinfo {author} {\bibfnamefont {R.}~\bibnamefont {Chétrite}},\ }\href
  {https://doi.org/10.1038/s42254-021-00349-8} {\bibfield  {journal} {\bibinfo
  {journal} {Nature Reviews Physics}\ }\textbf {\bibinfo {volume} {3}},\
  \bibinfo {pages} {534} (\bibinfo {year} {2021})}\BibitemShut {NoStop}%
\bibitem [{\citenamefont {Kumar}\ \emph {et~al.}(2022)\citenamefont {Kumar},
  \citenamefont {Chétrite},\ and\ \citenamefont {Bechhoefer}}]{kumar2022}%
  \BibitemOpen
  \bibfield  {author} {\bibinfo {author} {\bibfnamefont {A.}~\bibnamefont
  {Kumar}}, \bibinfo {author} {\bibfnamefont {R.}~\bibnamefont {Chétrite}},\
  and\ \bibinfo {author} {\bibfnamefont {J.}~\bibnamefont {Bechhoefer}},\
  }\href {https://www.pnas.org/doi/abs/10.1073/pnas.2118484119} {\bibfield
  {journal} {\bibinfo  {journal} {PNAS}\ }\textbf {\bibinfo {volume} {119}},\
  \bibinfo {pages} {e2118484119} (\bibinfo {year} {2022})}\BibitemShut
  {NoStop}%
\bibitem [{\citenamefont {Walker}\ and\ \citenamefont
  {Vucelja}()}]{walker2023mpemba}%
  \BibitemOpen
  \bibfield  {author} {\bibinfo {author} {\bibfnamefont {M.~R.}\ \bibnamefont
  {Walker}}\ and\ \bibinfo {author} {\bibfnamefont {M.}~\bibnamefont
  {Vucelja}},\ }\href@noop {} {}\Eprint {https://arxiv.org/abs/2212.07496}
  {arXiv:2212.07496} \BibitemShut {NoStop}%
\bibitem [{\citenamefont {Walker}\ \emph {et~al.}()\citenamefont {Walker},
  \citenamefont {Bera},\ and\ \citenamefont {Vucelja}}]{walker2023optimal}%
  \BibitemOpen
  \bibfield  {author} {\bibinfo {author} {\bibfnamefont {M.~R.}\ \bibnamefont
  {Walker}}, \bibinfo {author} {\bibfnamefont {S.}~\bibnamefont {Bera}},\ and\
  \bibinfo {author} {\bibfnamefont {M.}~\bibnamefont {Vucelja}},\ }\href@noop
  {} {}\Eprint {https://arxiv.org/abs/2307.16103} {arXiv:2307.16103}
  \BibitemShut {NoStop}%
\bibitem [{\citenamefont {Bera}\ \emph {et~al.}()\citenamefont {Bera},
  \citenamefont {Walker},\ and\ \citenamefont {Vucelja}}]{bera2023effect}%
  \BibitemOpen
  \bibfield  {author} {\bibinfo {author} {\bibfnamefont {S.}~\bibnamefont
  {Bera}}, \bibinfo {author} {\bibfnamefont {M.~R.}\ \bibnamefont {Walker}},\
  and\ \bibinfo {author} {\bibfnamefont {M.}~\bibnamefont {Vucelja}},\
  }\href@noop {} {}\Eprint {https://arxiv.org/abs/2308.04557}
  {arXiv:2308.04557} \BibitemShut {NoStop}%
\bibitem [{\citenamefont {Nava}\ and\ \citenamefont {Fabrizio}(2019)}]{nf-19}%
  \BibitemOpen
  \bibfield  {author} {\bibinfo {author} {\bibfnamefont {A.}~\bibnamefont
  {Nava}}\ and\ \bibinfo {author} {\bibfnamefont {M.}~\bibnamefont
  {Fabrizio}},\ }\href {https://doi.org/10.1103/PhysRevB.100.125102} {\bibfield
   {journal} {\bibinfo  {journal} {Phys. Rev. B}\ }\textbf {\bibinfo {volume}
  {100}},\ \bibinfo {pages} {125102} (\bibinfo {year} {2019})}\BibitemShut
  {NoStop}%
\bibitem [{\citenamefont {Chatterjee}\ \emph {et~al.}(2024)\citenamefont
  {Chatterjee}, \citenamefont {Takada},\ and\ \citenamefont
  {Hayakawa}}]{chatterjee2023multiple}%
  \BibitemOpen
  \bibfield  {author} {\bibinfo {author} {\bibfnamefont {A.~K.}\ \bibnamefont
  {Chatterjee}}, \bibinfo {author} {\bibfnamefont {S.}~\bibnamefont {Takada}},\
  and\ \bibinfo {author} {\bibfnamefont {H.}~\bibnamefont {Hayakawa}},\ }\href
  {https://doi.org/10.1103/PhysRevA.110.022213} {\bibfield  {journal} {\bibinfo
   {journal} {Phys. Rev. A}\ }\textbf {\bibinfo {volume} {110}},\ \bibinfo
  {pages} {022213} (\bibinfo {year} {2024})}\BibitemShut {NoStop}%
\bibitem [{\citenamefont {Chatterjee}\ \emph {et~al.}(2023)\citenamefont
  {Chatterjee}, \citenamefont {Takada},\ and\ \citenamefont
  {Hayakawa}}]{chatterjee2023}%
  \BibitemOpen
  \bibfield  {author} {\bibinfo {author} {\bibfnamefont {A.~K.}\ \bibnamefont
  {Chatterjee}}, \bibinfo {author} {\bibfnamefont {S.}~\bibnamefont {Takada}},\
  and\ \bibinfo {author} {\bibfnamefont {H.}~\bibnamefont {Hayakawa}},\ }\href
  {https://doi.org/10.1103/PhysRevLett.131.080402} {\bibfield  {journal}
  {\bibinfo  {journal} {Phys. Rev. Lett.}\ }\textbf {\bibinfo {volume} {131}},\
  \bibinfo {pages} {080402} (\bibinfo {year} {2023})}\BibitemShut {NoStop}%
\bibitem [{\citenamefont {Kochsiek}\ \emph {et~al.}(2022)\citenamefont
  {Kochsiek}, \citenamefont {Carollo},\ and\ \citenamefont
  {Lesanovsky}}]{kochsiek2022}%
  \BibitemOpen
  \bibfield  {author} {\bibinfo {author} {\bibfnamefont {S.}~\bibnamefont
  {Kochsiek}}, \bibinfo {author} {\bibfnamefont {F.}~\bibnamefont {Carollo}},\
  and\ \bibinfo {author} {\bibfnamefont {I.}~\bibnamefont {Lesanovsky}},\
  }\href {https://doi.org/10.1103/PhysRevA.106.012207} {\bibfield  {journal}
  {\bibinfo  {journal} {Phys. Rev. A}\ }\textbf {\bibinfo {volume} {106}},\
  \bibinfo {pages} {012207} (\bibinfo {year} {2022})}\BibitemShut {NoStop}%
\bibitem [{\citenamefont {Carollo}\ \emph {et~al.}(2021)\citenamefont
  {Carollo}, \citenamefont {Lasanta},\ and\ \citenamefont
  {Lesanovsky}}]{carollo2021}%
  \BibitemOpen
  \bibfield  {author} {\bibinfo {author} {\bibfnamefont {F.}~\bibnamefont
  {Carollo}}, \bibinfo {author} {\bibfnamefont {A.}~\bibnamefont {Lasanta}},\
  and\ \bibinfo {author} {\bibfnamefont {I.}~\bibnamefont {Lesanovsky}},\
  }\href {https://doi.org/10.1103/PhysRevLett.127.060401} {\bibfield  {journal}
  {\bibinfo  {journal} {Phys. Rev. Lett.}\ }\textbf {\bibinfo {volume} {127}},\
  \bibinfo {pages} {060401} (\bibinfo {year} {2021})}\BibitemShut {NoStop}%
\bibitem [{\citenamefont {Ivander}\ \emph {et~al.}(2023)\citenamefont
  {Ivander}, \citenamefont {Anto-Sztrikacs},\ and\ \citenamefont
  {Segal}}]{ias-23}%
  \BibitemOpen
  \bibfield  {author} {\bibinfo {author} {\bibfnamefont {F.}~\bibnamefont
  {Ivander}}, \bibinfo {author} {\bibfnamefont {N.}~\bibnamefont
  {Anto-Sztrikacs}},\ and\ \bibinfo {author} {\bibfnamefont {D.}~\bibnamefont
  {Segal}},\ }\href {https://doi.org/10.1103/PhysRevE.108.014130} {\bibfield
  {journal} {\bibinfo  {journal} {Phys. Rev. E}\ }\textbf {\bibinfo {volume}
  {108}},\ \bibinfo {pages} {014130} (\bibinfo {year} {2023})}\BibitemShut
  {NoStop}%
\bibitem [{\citenamefont {Shapira}\ \emph {et~al.}(2024)\citenamefont
  {Shapira}, \citenamefont {Shapira}, \citenamefont {Markov}, \citenamefont
  {Teza}, \citenamefont {Akerman}, \citenamefont {Raz},\ and\ \citenamefont
  {Ozeri}}]{shapira2024}%
  \BibitemOpen
  \bibfield  {author} {\bibinfo {author} {\bibfnamefont {S.~A.}\ \bibnamefont
  {Shapira}}, \bibinfo {author} {\bibfnamefont {Y.}~\bibnamefont {Shapira}},
  \bibinfo {author} {\bibfnamefont {J.}~\bibnamefont {Markov}}, \bibinfo
  {author} {\bibfnamefont {G.}~\bibnamefont {Teza}}, \bibinfo {author}
  {\bibfnamefont {N.}~\bibnamefont {Akerman}}, \bibinfo {author} {\bibfnamefont
  {O.}~\bibnamefont {Raz}},\ and\ \bibinfo {author} {\bibfnamefont
  {R.}~\bibnamefont {Ozeri}},\ }\href {https://arxiv.org/abs/2401.05830}
  {\bibfield  {journal} {\bibinfo  {journal} {arXiv:2401.05830}\ } (\bibinfo
  {year} {2024})}\BibitemShut {NoStop}%
\bibitem [{\citenamefont {Strachan}\ \emph {et~al.}()\citenamefont {Strachan},
  \citenamefont {Purkayastha},\ and\ \citenamefont
  {Clark}}]{strachan2024nonmarkovian}%
  \BibitemOpen
  \bibfield  {author} {\bibinfo {author} {\bibfnamefont {D.~J.}\ \bibnamefont
  {Strachan}}, \bibinfo {author} {\bibfnamefont {A.}~\bibnamefont
  {Purkayastha}},\ and\ \bibinfo {author} {\bibfnamefont {S.~R.}\ \bibnamefont
  {Clark}},\ }\href@noop {} {}\Eprint {https://arxiv.org/abs/2402.05756}
  {arXiv:2402.05756} \BibitemShut {NoStop}%
\bibitem [{\citenamefont {Zhang}\ \emph {et~al.}()\citenamefont {Zhang},
  \citenamefont {Xia}, \citenamefont {Wu}, \citenamefont {Chen}, \citenamefont
  {Zhang}, \citenamefont {Xie}, \citenamefont {Su}, \citenamefont {Wu},
  \citenamefont {Qiu}, \citenamefont {xing Chen}, \citenamefont {Li},
  \citenamefont {Jing},\ and\ \citenamefont {Zhou}}]{zhang2024observation}%
  \BibitemOpen
  \bibfield  {author} {\bibinfo {author} {\bibfnamefont {J.}~\bibnamefont
  {Zhang}}, \bibinfo {author} {\bibfnamefont {G.}~\bibnamefont {Xia}}, \bibinfo
  {author} {\bibfnamefont {C.-W.}\ \bibnamefont {Wu}}, \bibinfo {author}
  {\bibfnamefont {T.}~\bibnamefont {Chen}}, \bibinfo {author} {\bibfnamefont
  {Q.}~\bibnamefont {Zhang}}, \bibinfo {author} {\bibfnamefont
  {Y.}~\bibnamefont {Xie}}, \bibinfo {author} {\bibfnamefont {W.-B.}\
  \bibnamefont {Su}}, \bibinfo {author} {\bibfnamefont {W.}~\bibnamefont {Wu}},
  \bibinfo {author} {\bibfnamefont {C.-W.}\ \bibnamefont {Qiu}}, \bibinfo
  {author} {\bibfnamefont {P.}~\bibnamefont {xing Chen}}, \bibinfo {author}
  {\bibfnamefont {W.}~\bibnamefont {Li}}, \bibinfo {author} {\bibfnamefont
  {H.}~\bibnamefont {Jing}},\ and\ \bibinfo {author} {\bibfnamefont {Y.-L.}\
  \bibnamefont {Zhou}},\ }\href@noop {} {}\Eprint
  {https://arxiv.org/abs/2401.15951} {arXiv:2401.15951} \BibitemShut {NoStop}%
\bibitem [{\citenamefont {Wang}\ and\ \citenamefont
  {Wang}(2024)}]{wang2024mpemba}%
  \BibitemOpen
  \bibfield  {author} {\bibinfo {author} {\bibfnamefont {X.}~\bibnamefont
  {Wang}}\ and\ \bibinfo {author} {\bibfnamefont {J.}~\bibnamefont {Wang}},\
  }\href {https://doi.org/10.1103/PhysRevResearch.6.033330} {\bibfield
  {journal} {\bibinfo  {journal} {Phys. Rev. Res.}\ }\textbf {\bibinfo {volume}
  {6}},\ \bibinfo {pages} {033330} (\bibinfo {year} {2024})}\BibitemShut
  {NoStop}%
\bibitem [{\citenamefont {Moroder}\ \emph {et~al.}(2024)\citenamefont
  {Moroder}, \citenamefont {Culhane}, \citenamefont {Zawadzki},\ and\
  \citenamefont {Goold}}]{moroder2024thermodynamics}%
  \BibitemOpen
  \bibfield  {author} {\bibinfo {author} {\bibfnamefont {M.}~\bibnamefont
  {Moroder}}, \bibinfo {author} {\bibfnamefont {O.}~\bibnamefont {Culhane}},
  \bibinfo {author} {\bibfnamefont {K.}~\bibnamefont {Zawadzki}},\ and\
  \bibinfo {author} {\bibfnamefont {J.}~\bibnamefont {Goold}},\ }\href
  {https://doi.org/10.1103/PhysRevLett.133.140404} {\bibfield  {journal}
  {\bibinfo  {journal} {Phys. Rev. Lett.}\ }\textbf {\bibinfo {volume} {133}},\
  \bibinfo {pages} {140404} (\bibinfo {year} {2024})}\BibitemShut {NoStop}%
\bibitem [{\citenamefont {Ares}\ \emph
  {et~al.}(2023{\natexlab{a}})\citenamefont {Ares}, \citenamefont {Murciano},\
  and\ \citenamefont {Calabrese}}]{ares2022entanglement}%
  \BibitemOpen
  \bibfield  {author} {\bibinfo {author} {\bibfnamefont {F.}~\bibnamefont
  {Ares}}, \bibinfo {author} {\bibfnamefont {S.}~\bibnamefont {Murciano}},\
  and\ \bibinfo {author} {\bibfnamefont {P.}~\bibnamefont {Calabrese}},\ }\href
  {https://doi.org/10.1038/s41467-023-37747-8} {\bibfield  {journal} {\bibinfo
  {journal} {Nat. Commun.}\ }\textbf {\bibinfo {volume} {14}},\ \bibinfo
  {pages} {2036} (\bibinfo {year} {2023}{\natexlab{a}})}\BibitemShut {NoStop}%
\bibitem [{\citenamefont {Ares}\ \emph
  {et~al.}(2023{\natexlab{b}})\citenamefont {Ares}, \citenamefont {Murciano},
  \citenamefont {Vernier},\ and\ \citenamefont {Calabrese}}]{ares2023}%
  \BibitemOpen
  \bibfield  {author} {\bibinfo {author} {\bibfnamefont {F.}~\bibnamefont
  {Ares}}, \bibinfo {author} {\bibfnamefont {S.}~\bibnamefont {Murciano}},
  \bibinfo {author} {\bibfnamefont {E.}~\bibnamefont {Vernier}},\ and\ \bibinfo
  {author} {\bibfnamefont {P.}~\bibnamefont {Calabrese}},\ }\href
  {https://doi.org/10.21468/SciPostPhys.15.3.089} {\bibfield  {journal}
  {\bibinfo  {journal} {SciPost Phys.}\ }\textbf {\bibinfo {volume} {15}},\
  \bibinfo {pages} {089} (\bibinfo {year} {2023}{\natexlab{b}})}\BibitemShut
  {NoStop}%
\bibitem [{\citenamefont {Murciano}\ \emph {et~al.}(2024)\citenamefont
  {Murciano}, \citenamefont {Ares}, \citenamefont {Klich},\ and\ \citenamefont
  {Calabrese}}]{Murciano_2024}%
  \BibitemOpen
  \bibfield  {author} {\bibinfo {author} {\bibfnamefont {S.}~\bibnamefont
  {Murciano}}, \bibinfo {author} {\bibfnamefont {F.}~\bibnamefont {Ares}},
  \bibinfo {author} {\bibfnamefont {I.}~\bibnamefont {Klich}},\ and\ \bibinfo
  {author} {\bibfnamefont {P.}~\bibnamefont {Calabrese}},\ }\href
  {https://doi.org/10.1088/1742-5468/ad17b4} {\bibfield  {journal} {\bibinfo
  {journal} {J. Stat. Mech.}\ }\textbf {\bibinfo {volume} {2024}},\ \bibinfo
  {pages} {013103} (\bibinfo {year} {2024})}\BibitemShut {NoStop}%
\bibitem [{\citenamefont {Yamashika}\ \emph {et~al.}(2024)\citenamefont
  {Yamashika}, \citenamefont {Ares},\ and\ \citenamefont
  {Calabrese}}]{yamashika2024entanglement}%
  \BibitemOpen
  \bibfield  {author} {\bibinfo {author} {\bibfnamefont {S.}~\bibnamefont
  {Yamashika}}, \bibinfo {author} {\bibfnamefont {F.}~\bibnamefont {Ares}},\
  and\ \bibinfo {author} {\bibfnamefont {P.}~\bibnamefont {Calabrese}},\ }\href
  {https://doi.org/10.1103/PhysRevB.110.085126} {\bibfield  {journal} {\bibinfo
   {journal} {Phys. Rev. B}\ }\textbf {\bibinfo {volume} {110}},\ \bibinfo
  {pages} {085126} (\bibinfo {year} {2024})}\BibitemShut {NoStop}%
\bibitem [{\citenamefont {Caceffo}\ \emph {et~al.}(2024)\citenamefont
  {Caceffo}, \citenamefont {Murciano},\ and\ \citenamefont
  {Alba}}]{caceffo2024entangled}%
  \BibitemOpen
  \bibfield  {author} {\bibinfo {author} {\bibfnamefont {F.}~\bibnamefont
  {Caceffo}}, \bibinfo {author} {\bibfnamefont {S.}~\bibnamefont {Murciano}},\
  and\ \bibinfo {author} {\bibfnamefont {V.}~\bibnamefont {Alba}},\ }\href
  {https://doi.org/10.1088/1742-5468/ad4537} {\bibfield  {journal} {\bibinfo
  {journal} {J. Stat. Mech.: Theor. Exp.}\ }\textbf {\bibinfo {volume}
  {2024}},\ \bibinfo {pages} {063103} (\bibinfo {year} {2024})}\BibitemShut
  {NoStop}%
\bibitem [{\citenamefont {Bertini}\ \emph {et~al.}(2024)\citenamefont
  {Bertini}, \citenamefont {Klobas}, \citenamefont {Collura}, \citenamefont
  {Calabrese},\ and\ \citenamefont {Rylands}}]{bertini2023dynamics}%
  \BibitemOpen
  \bibfield  {author} {\bibinfo {author} {\bibfnamefont {B.}~\bibnamefont
  {Bertini}}, \bibinfo {author} {\bibfnamefont {K.}~\bibnamefont {Klobas}},
  \bibinfo {author} {\bibfnamefont {M.}~\bibnamefont {Collura}}, \bibinfo
  {author} {\bibfnamefont {P.}~\bibnamefont {Calabrese}},\ and\ \bibinfo
  {author} {\bibfnamefont {C.}~\bibnamefont {Rylands}},\ }\href
  {https://doi.org/10.1103/PhysRevB.109.184312} {\bibfield  {journal} {\bibinfo
   {journal} {Phys. Rev. B}\ }\textbf {\bibinfo {volume} {109}},\ \bibinfo
  {pages} {184312} (\bibinfo {year} {2024})}\BibitemShut {NoStop}%
\bibitem [{\citenamefont {Ferro}\ \emph {et~al.}(2024)\citenamefont {Ferro},
  \citenamefont {Ares},\ and\ \citenamefont {Calabrese}}]{ferro2024}%
  \BibitemOpen
  \bibfield  {author} {\bibinfo {author} {\bibfnamefont {F.}~\bibnamefont
  {Ferro}}, \bibinfo {author} {\bibfnamefont {F.}~\bibnamefont {Ares}},\ and\
  \bibinfo {author} {\bibfnamefont {P.}~\bibnamefont {Calabrese}},\ }\href
  {https://dx.doi.org/10.1088/1742-5468/ad138f} {\bibfield  {journal} {\bibinfo
   {journal} {J. Stat. Mech.}\ }\textbf {\bibinfo {volume} {2024}},\ \bibinfo
  {pages} {023101} (\bibinfo {year} {2024})}\BibitemShut {NoStop}%
\bibitem [{\citenamefont {Rylands}\ \emph {et~al.}(2024)\citenamefont
  {Rylands}, \citenamefont {Klobas}, \citenamefont {Ares}, \citenamefont
  {Calabrese}, \citenamefont {Murciano},\ and\ \citenamefont
  {Bertini}}]{rylands2023microscopic}%
  \BibitemOpen
  \bibfield  {author} {\bibinfo {author} {\bibfnamefont {C.}~\bibnamefont
  {Rylands}}, \bibinfo {author} {\bibfnamefont {K.}~\bibnamefont {Klobas}},
  \bibinfo {author} {\bibfnamefont {F.}~\bibnamefont {Ares}}, \bibinfo {author}
  {\bibfnamefont {P.}~\bibnamefont {Calabrese}}, \bibinfo {author}
  {\bibfnamefont {S.}~\bibnamefont {Murciano}},\ and\ \bibinfo {author}
  {\bibfnamefont {B.}~\bibnamefont {Bertini}},\ }\href
  {https://doi.org/10.1103/PhysRevLett.133.010401} {\bibfield  {journal}
  {\bibinfo  {journal} {Phys. Rev. Lett.}\ }\textbf {\bibinfo {volume} {133}},\
  \bibinfo {pages} {010401} (\bibinfo {year} {2024})}\BibitemShut {NoStop}%
\bibitem [{\citenamefont {Chalas}\ \emph {et~al.}(2024)\citenamefont {Chalas},
  \citenamefont {Ares}, \citenamefont {Rylands},\ and\ \citenamefont
  {Calabrese}}]{chalas2024multiple}%
  \BibitemOpen
  \bibfield  {author} {\bibinfo {author} {\bibfnamefont {K.}~\bibnamefont
  {Chalas}}, \bibinfo {author} {\bibfnamefont {F.}~\bibnamefont {Ares}},
  \bibinfo {author} {\bibfnamefont {C.}~\bibnamefont {Rylands}},\ and\ \bibinfo
  {author} {\bibfnamefont {P.}~\bibnamefont {Calabrese}},\ }\href
  {https://doi.org/10.1088/1742-5468/ad769c} {\bibfield  {journal} {\bibinfo
  {journal} {J. Stat. Mech.: Theor. Exp.}\ }\textbf {\bibinfo {volume}
  {2024}},\ \bibinfo {pages} {103101} (\bibinfo {year} {2024})}\BibitemShut
  {NoStop}%
\bibitem [{\citenamefont {Ares}\ \emph {et~al.}()\citenamefont {Ares},
  \citenamefont {Vitale},\ and\ \citenamefont {Murciano}}]{ares2024quantum}%
  \BibitemOpen
  \bibfield  {author} {\bibinfo {author} {\bibfnamefont {F.}~\bibnamefont
  {Ares}}, \bibinfo {author} {\bibfnamefont {V.}~\bibnamefont {Vitale}},\ and\
  \bibinfo {author} {\bibfnamefont {S.}~\bibnamefont {Murciano}},\ }\href@noop
  {} {}\Eprint {https://arxiv.org/abs/2405.08913} {arXiv:2405.08913}
  \BibitemShut {NoStop}%
\bibitem [{\citenamefont {Joshi}\ \emph {et~al.}(2024)\citenamefont {Joshi},
  \citenamefont {Franke}, \citenamefont {Rath}, \citenamefont {Ares},
  \citenamefont {Murciano}, \citenamefont {Kranzl}, \citenamefont {Blatt},
  \citenamefont {Zoller}, \citenamefont {Vermersch}, \citenamefont {Calabrese},
  \citenamefont {Roos},\ and\ \citenamefont {Joshi}}]{joshi2024}%
  \BibitemOpen
  \bibfield  {author} {\bibinfo {author} {\bibfnamefont {L.~K.}\ \bibnamefont
  {Joshi}}, \bibinfo {author} {\bibfnamefont {J.}~\bibnamefont {Franke}},
  \bibinfo {author} {\bibfnamefont {A.}~\bibnamefont {Rath}}, \bibinfo {author}
  {\bibfnamefont {F.}~\bibnamefont {Ares}}, \bibinfo {author} {\bibfnamefont
  {S.}~\bibnamefont {Murciano}}, \bibinfo {author} {\bibfnamefont
  {F.}~\bibnamefont {Kranzl}}, \bibinfo {author} {\bibfnamefont
  {R.}~\bibnamefont {Blatt}}, \bibinfo {author} {\bibfnamefont
  {P.}~\bibnamefont {Zoller}}, \bibinfo {author} {\bibfnamefont
  {B.}~\bibnamefont {Vermersch}}, \bibinfo {author} {\bibfnamefont
  {P.}~\bibnamefont {Calabrese}}, \bibinfo {author} {\bibfnamefont {C.~F.}\
  \bibnamefont {Roos}},\ and\ \bibinfo {author} {\bibfnamefont {M.~K.}\
  \bibnamefont {Joshi}},\ }\href
  {https://doi.org/10.1103/PhysRevLett.133.010402} {\bibfield  {journal}
  {\bibinfo  {journal} {Phys. Rev. Lett.}\ }\textbf {\bibinfo {volume} {133}},\
  \bibinfo {pages} {010402} (\bibinfo {year} {2024})}\BibitemShut {NoStop}%
\bibitem [{\citenamefont {Nahum}\ \emph {et~al.}(2017)\citenamefont {Nahum},
  \citenamefont {Ruhman}, \citenamefont {Vijay},\ and\ \citenamefont
  {Haah}}]{nahum2017}%
  \BibitemOpen
  \bibfield  {author} {\bibinfo {author} {\bibfnamefont {A.}~\bibnamefont
  {Nahum}}, \bibinfo {author} {\bibfnamefont {J.}~\bibnamefont {Ruhman}},
  \bibinfo {author} {\bibfnamefont {S.}~\bibnamefont {Vijay}},\ and\ \bibinfo
  {author} {\bibfnamefont {J.}~\bibnamefont {Haah}},\ }\href
  {https://doi.org/10.1103/PhysRevX.7.031016} {\bibfield  {journal} {\bibinfo
  {journal} {Phys. Rev. X}\ }\textbf {\bibinfo {volume} {7}},\ \bibinfo {pages}
  {031016} (\bibinfo {year} {2017})}\BibitemShut {NoStop}%
\bibitem [{\citenamefont {Nahum}\ \emph {et~al.}(2018)\citenamefont {Nahum},
  \citenamefont {Vijay},\ and\ \citenamefont {Haah}}]{nahum2018}%
  \BibitemOpen
  \bibfield  {author} {\bibinfo {author} {\bibfnamefont {A.}~\bibnamefont
  {Nahum}}, \bibinfo {author} {\bibfnamefont {S.}~\bibnamefont {Vijay}},\ and\
  \bibinfo {author} {\bibfnamefont {J.}~\bibnamefont {Haah}},\ }\href
  {https://doi.org/10.1103/PhysRevX.8.021014} {\bibfield  {journal} {\bibinfo
  {journal} {Phys. Rev. X}\ }\textbf {\bibinfo {volume} {8}},\ \bibinfo {pages}
  {021014} (\bibinfo {year} {2018})}\BibitemShut {NoStop}%
\bibitem [{\citenamefont {Khemani}\ \emph
  {et~al.}(2018{\natexlab{a}})\citenamefont {Khemani}, \citenamefont
  {Vishwanath},\ and\ \citenamefont {Huse}}]{khemani2018}%
  \BibitemOpen
  \bibfield  {author} {\bibinfo {author} {\bibfnamefont {V.}~\bibnamefont
  {Khemani}}, \bibinfo {author} {\bibfnamefont {A.}~\bibnamefont
  {Vishwanath}},\ and\ \bibinfo {author} {\bibfnamefont {D.~A.}\ \bibnamefont
  {Huse}},\ }\href {https://doi.org/10.1103/PhysRevX.8.031057} {\bibfield
  {journal} {\bibinfo  {journal} {Phys. Rev. X}\ }\textbf {\bibinfo {volume}
  {8}},\ \bibinfo {pages} {031057} (\bibinfo {year}
  {2018}{\natexlab{a}})}\BibitemShut {NoStop}%
\bibitem [{\citenamefont {von Keyserlingk}\ \emph {et~al.}(2018)\citenamefont
  {von Keyserlingk}, \citenamefont {Rakovszky}, \citenamefont {Pollmann},\ and\
  \citenamefont {Sondhi}}]{keyserlingk2018}%
  \BibitemOpen
  \bibfield  {author} {\bibinfo {author} {\bibfnamefont {C.~W.}\ \bibnamefont
  {von Keyserlingk}}, \bibinfo {author} {\bibfnamefont {T.}~\bibnamefont
  {Rakovszky}}, \bibinfo {author} {\bibfnamefont {F.}~\bibnamefont
  {Pollmann}},\ and\ \bibinfo {author} {\bibfnamefont {S.~L.}\ \bibnamefont
  {Sondhi}},\ }\href {https://doi.org/10.1103/PhysRevX.8.021013} {\bibfield
  {journal} {\bibinfo  {journal} {Phys. Rev. X}\ }\textbf {\bibinfo {volume}
  {8}},\ \bibinfo {pages} {021013} (\bibinfo {year} {2018})}\BibitemShut
  {NoStop}%
\bibitem [{\citenamefont {Fisher}\ \emph {et~al.}(2023)\citenamefont {Fisher},
  \citenamefont {Khemani}, \citenamefont {Nahum},\ and\ \citenamefont
  {Vijay}}]{fisher2023}%
  \BibitemOpen
  \bibfield  {author} {\bibinfo {author} {\bibfnamefont {M.~P.}\ \bibnamefont
  {Fisher}}, \bibinfo {author} {\bibfnamefont {V.}~\bibnamefont {Khemani}},
  \bibinfo {author} {\bibfnamefont {A.}~\bibnamefont {Nahum}},\ and\ \bibinfo
  {author} {\bibfnamefont {S.}~\bibnamefont {Vijay}},\ }\href
  {https://doi.org/10.1146/annurev-conmatphys-031720-030658} {\bibfield
  {journal} {\bibinfo  {journal} {Ann. Rev. Cond. Matter Phys.}\ }\textbf
  {\bibinfo {volume} {14}},\ \bibinfo {pages} {335} (\bibinfo {year}
  {2023})}\BibitemShut {NoStop}%
\bibitem [{\citenamefont {Žnidarič}(2020)}]{nidari2020}%
  \BibitemOpen
  \bibfield  {author} {\bibinfo {author} {\bibfnamefont {M.}~\bibnamefont
  {Žnidarič}},\ }\href {http://dx.doi.org/10.1038/s42005-020-0366-7}
  {\bibfield  {journal} {\bibinfo  {journal} {Comm. Phys.}\ }\textbf {\bibinfo
  {volume} {3}},\ \bibinfo {pages} {100} (\bibinfo {year} {2020})}\BibitemShut
  {NoStop}%
\bibitem [{\citenamefont {Sierant}\ \emph {et~al.}(2023)\citenamefont
  {Sierant}, \citenamefont {Schir\`o}, \citenamefont {Lewenstein},\ and\
  \citenamefont {Turkeshi}}]{sierant2023}%
  \BibitemOpen
  \bibfield  {author} {\bibinfo {author} {\bibfnamefont {P.}~\bibnamefont
  {Sierant}}, \bibinfo {author} {\bibfnamefont {M.}~\bibnamefont {Schir\`o}},
  \bibinfo {author} {\bibfnamefont {M.}~\bibnamefont {Lewenstein}},\ and\
  \bibinfo {author} {\bibfnamefont {X.}~\bibnamefont {Turkeshi}},\ }\href
  {https://doi.org/10.1103/PhysRevLett.131.230403} {\bibfield  {journal}
  {\bibinfo  {journal} {Phys. Rev. Lett.}\ }\textbf {\bibinfo {volume} {131}},\
  \bibinfo {pages} {230403} (\bibinfo {year} {2023})}\BibitemShut {NoStop}%
\bibitem [{\citenamefont {Richter}\ \emph {et~al.}(2023)\citenamefont
  {Richter}, \citenamefont {Lunt},\ and\ \citenamefont {Pal}}]{richter2023}%
  \BibitemOpen
  \bibfield  {author} {\bibinfo {author} {\bibfnamefont {J.}~\bibnamefont
  {Richter}}, \bibinfo {author} {\bibfnamefont {O.}~\bibnamefont {Lunt}},\ and\
  \bibinfo {author} {\bibfnamefont {A.}~\bibnamefont {Pal}},\ }\href
  {https://doi.org/10.1103/PhysRevResearch.5.L012031} {\bibfield  {journal}
  {\bibinfo  {journal} {Phys. Rev. Res.}\ }\textbf {\bibinfo {volume} {5}},\
  \bibinfo {pages} {L012031} (\bibinfo {year} {2023})}\BibitemShut {NoStop}%
\bibitem [{\citenamefont {Jonay}\ \emph {et~al.}(2024)\citenamefont {Jonay},
  \citenamefont {Rodriguez-Nieva},\ and\ \citenamefont {Khemani}}]{jonay2024}%
  \BibitemOpen
  \bibfield  {author} {\bibinfo {author} {\bibfnamefont {C.}~\bibnamefont
  {Jonay}}, \bibinfo {author} {\bibfnamefont {J.~F.}\ \bibnamefont
  {Rodriguez-Nieva}},\ and\ \bibinfo {author} {\bibfnamefont {V.}~\bibnamefont
  {Khemani}},\ }\href {https://doi.org/10.1103/PhysRevB.109.024311} {\bibfield
  {journal} {\bibinfo  {journal} {Phys. Rev. B}\ }\textbf {\bibinfo {volume}
  {109}},\ \bibinfo {pages} {024311} (\bibinfo {year} {2024})}\BibitemShut
  {NoStop}%
\bibitem [{\citenamefont {Chan}\ \emph
  {et~al.}(2018{\natexlab{a}})\citenamefont {Chan}, \citenamefont {De~Luca},\
  and\ \citenamefont {Chalker}}]{chan2018}%
  \BibitemOpen
  \bibfield  {author} {\bibinfo {author} {\bibfnamefont {A.}~\bibnamefont
  {Chan}}, \bibinfo {author} {\bibfnamefont {A.}~\bibnamefont {De~Luca}},\ and\
  \bibinfo {author} {\bibfnamefont {J.~T.}\ \bibnamefont {Chalker}},\ }\href
  {https://doi.org/10.1103/PhysRevX.8.041019} {\bibfield  {journal} {\bibinfo
  {journal} {Phys. Rev. X}\ }\textbf {\bibinfo {volume} {8}},\ \bibinfo {pages}
  {041019} (\bibinfo {year} {2018}{\natexlab{a}})}\BibitemShut {NoStop}%
\bibitem [{\citenamefont {Chan}\ \emph
  {et~al.}(2018{\natexlab{b}})\citenamefont {Chan}, \citenamefont {De~Luca},\
  and\ \citenamefont {Chalker}}]{chan20182}%
  \BibitemOpen
  \bibfield  {author} {\bibinfo {author} {\bibfnamefont {A.}~\bibnamefont
  {Chan}}, \bibinfo {author} {\bibfnamefont {A.}~\bibnamefont {De~Luca}},\ and\
  \bibinfo {author} {\bibfnamefont {J.~T.}\ \bibnamefont {Chalker}},\ }\href
  {https://doi.org/10.1103/PhysRevLett.121.060601} {\bibfield  {journal}
  {\bibinfo  {journal} {Phys. Rev. Lett.}\ }\textbf {\bibinfo {volume} {121}},\
  \bibinfo {pages} {060601} (\bibinfo {year} {2018}{\natexlab{b}})}\BibitemShut
  {NoStop}%
\bibitem [{\citenamefont {Chan}\ \emph {et~al.}(2019)\citenamefont {Chan},
  \citenamefont {De~Luca},\ and\ \citenamefont {Chalker}}]{chan2019}%
  \BibitemOpen
  \bibfield  {author} {\bibinfo {author} {\bibfnamefont {A.}~\bibnamefont
  {Chan}}, \bibinfo {author} {\bibfnamefont {A.}~\bibnamefont {De~Luca}},\ and\
  \bibinfo {author} {\bibfnamefont {J.~T.}\ \bibnamefont {Chalker}},\ }\href
  {https://doi.org/10.1103/PhysRevLett.122.220601} {\bibfield  {journal}
  {\bibinfo  {journal} {Phys. Rev. Lett.}\ }\textbf {\bibinfo {volume} {122}},\
  \bibinfo {pages} {220601} (\bibinfo {year} {2019})}\BibitemShut {NoStop}%
\bibitem [{\citenamefont {Friedman}\ \emph {et~al.}(2019)\citenamefont
  {Friedman}, \citenamefont {Chan}, \citenamefont {De~Luca},\ and\
  \citenamefont {Chalker}}]{chan20192}%
  \BibitemOpen
  \bibfield  {author} {\bibinfo {author} {\bibfnamefont {A.~J.}\ \bibnamefont
  {Friedman}}, \bibinfo {author} {\bibfnamefont {A.}~\bibnamefont {Chan}},
  \bibinfo {author} {\bibfnamefont {A.}~\bibnamefont {De~Luca}},\ and\ \bibinfo
  {author} {\bibfnamefont {J.~T.}\ \bibnamefont {Chalker}},\ }\href
  {https://doi.org/10.1103/PhysRevLett.123.210603} {\bibfield  {journal}
  {\bibinfo  {journal} {Phys. Rev. Lett.}\ }\textbf {\bibinfo {volume} {123}},\
  \bibinfo {pages} {210603} (\bibinfo {year} {2019})}\BibitemShut {NoStop}%
\bibitem [{\citenamefont {Chan}\ \emph {et~al.}(2021)\citenamefont {Chan},
  \citenamefont {De~Luca},\ and\ \citenamefont {Chalker}}]{chan2021}%
  \BibitemOpen
  \bibfield  {author} {\bibinfo {author} {\bibfnamefont {A.}~\bibnamefont
  {Chan}}, \bibinfo {author} {\bibfnamefont {A.}~\bibnamefont {De~Luca}},\ and\
  \bibinfo {author} {\bibfnamefont {J.~T.}\ \bibnamefont {Chalker}},\ }\href
  {https://doi.org/10.1103/PhysRevResearch.3.023118} {\bibfield  {journal}
  {\bibinfo  {journal} {Phys. Rev. Res.}\ }\textbf {\bibinfo {volume} {3}},\
  \bibinfo {pages} {023118} (\bibinfo {year} {2021})}\BibitemShut {NoStop}%
\bibitem [{\citenamefont {Chan}\ \emph {et~al.}(2022)\citenamefont {Chan},
  \citenamefont {Shivam}, \citenamefont {Huse},\ and\ \citenamefont
  {De~Luca}}]{Chan2022}%
  \BibitemOpen
  \bibfield  {author} {\bibinfo {author} {\bibfnamefont {A.}~\bibnamefont
  {Chan}}, \bibinfo {author} {\bibfnamefont {S.}~\bibnamefont {Shivam}},
  \bibinfo {author} {\bibfnamefont {D.~A.}\ \bibnamefont {Huse}},\ and\
  \bibinfo {author} {\bibfnamefont {A.}~\bibnamefont {De~Luca}},\ }\href
  {http://dx.doi.org/10.1038/s41467-022-34318-1} {\bibfield  {journal}
  {\bibinfo  {journal} {Nature Commun.}\ }\textbf {\bibinfo {volume} {13}},\
  \bibinfo {pages} {7484} (\bibinfo {year} {2022})}\BibitemShut {NoStop}%
\bibitem [{\citenamefont {Rakovszky}\ \emph {et~al.}(2019)\citenamefont
  {Rakovszky}, \citenamefont {Pollmann},\ and\ \citenamefont {von
  Keyserlingk}}]{rakovszky2019}%
  \BibitemOpen
  \bibfield  {author} {\bibinfo {author} {\bibfnamefont {T.}~\bibnamefont
  {Rakovszky}}, \bibinfo {author} {\bibfnamefont {F.}~\bibnamefont
  {Pollmann}},\ and\ \bibinfo {author} {\bibfnamefont {C.~W.}\ \bibnamefont
  {von Keyserlingk}},\ }\href {https://doi.org/10.1103/PhysRevLett.122.250602}
  {\bibfield  {journal} {\bibinfo  {journal} {Phys. Rev. Lett.}\ }\textbf
  {\bibinfo {volume} {122}},\ \bibinfo {pages} {250602} (\bibinfo {year}
  {2019})}\BibitemShut {NoStop}%
\bibitem [{\citenamefont {Zhou}\ and\ \citenamefont {Nahum}(2019)}]{zhou2019}%
  \BibitemOpen
  \bibfield  {author} {\bibinfo {author} {\bibfnamefont {T.}~\bibnamefont
  {Zhou}}\ and\ \bibinfo {author} {\bibfnamefont {A.}~\bibnamefont {Nahum}},\
  }\href {https://doi.org/10.1103/PhysRevB.99.174205} {\bibfield  {journal}
  {\bibinfo  {journal} {Phys. Rev. B}\ }\textbf {\bibinfo {volume} {99}},\
  \bibinfo {pages} {174205} (\bibinfo {year} {2019})}\BibitemShut {NoStop}%
\bibitem [{\citenamefont {Rakovszky}\ \emph {et~al.}(2018)\citenamefont
  {Rakovszky}, \citenamefont {Pollmann},\ and\ \citenamefont {von
  Keyserlingk}}]{PhysRevX.8.031058}%
  \BibitemOpen
  \bibfield  {author} {\bibinfo {author} {\bibfnamefont {T.}~\bibnamefont
  {Rakovszky}}, \bibinfo {author} {\bibfnamefont {F.}~\bibnamefont
  {Pollmann}},\ and\ \bibinfo {author} {\bibfnamefont {C.~W.}\ \bibnamefont
  {von Keyserlingk}},\ }\href {https://doi.org/10.1103/PhysRevX.8.031058}
  {\bibfield  {journal} {\bibinfo  {journal} {Phys. Rev. X}\ }\textbf {\bibinfo
  {volume} {8}},\ \bibinfo {pages} {031058} (\bibinfo {year}
  {2018})}\BibitemShut {NoStop}%
\bibitem [{Note1()}]{Note1}%
  \BibitemOpen
  \bibinfo {note} {For the sake of presentation, a detailed survay of the
  technicalities is presented in~\cite {supmat}.}\BibitemShut {Stop}%
\bibitem [{Note2()}]{Note2}%
  \BibitemOpen
  \bibinfo {note} {Here $U_{N,N+1}\equiv U_{N,1}$ ($U_{N,N+1}\equiv I_1\otimes
  I_N$) for periodic (open) boundary conditions.}\BibitemShut {Stop}%
\bibitem [{\citenamefont {Fossati}\ \emph {et~al.}(2024)\citenamefont
  {Fossati}, \citenamefont {Ares}, \citenamefont {Dubail},\ and\ \citenamefont
  {Calabrese}}]{fossati2024entanglement}%
  \BibitemOpen
  \bibfield  {author} {\bibinfo {author} {\bibfnamefont {M.}~\bibnamefont
  {Fossati}}, \bibinfo {author} {\bibfnamefont {F.}~\bibnamefont {Ares}},
  \bibinfo {author} {\bibfnamefont {J.}~\bibnamefont {Dubail}},\ and\ \bibinfo
  {author} {\bibfnamefont {P.}~\bibnamefont {Calabrese}},\ }\href
  {https://doi.org/10.1007/JHEP05(2024)059} {\bibfield  {journal} {\bibinfo
  {journal} {J. High Energy Phys.}\ }\textbf {\bibinfo {volume} {2024}},\
  \bibinfo {pages} {059}}\BibitemShut {NoStop}%
\bibitem [{\citenamefont {Chen}\ and\ \citenamefont {Chen}(2024)}]{chen2023}%
  \BibitemOpen
  \bibfield  {author} {\bibinfo {author} {\bibfnamefont {M.}~\bibnamefont
  {Chen}}\ and\ \bibinfo {author} {\bibfnamefont {H.-H.}\ \bibnamefont
  {Chen}},\ }\href {https://doi.org/10.1103/PhysRevD.109.065009} {\bibfield
  {journal} {\bibinfo  {journal} {Phys. Rev. D}\ }\textbf {\bibinfo {volume}
  {109}},\ \bibinfo {pages} {065009} (\bibinfo {year} {2024})}\BibitemShut
  {NoStop}%
\bibitem [{\citenamefont {Capizzi}\ and\ \citenamefont
  {Mazzoni}(2023)}]{capizzi2023}%
  \BibitemOpen
  \bibfield  {author} {\bibinfo {author} {\bibfnamefont {L.}~\bibnamefont
  {Capizzi}}\ and\ \bibinfo {author} {\bibfnamefont {M.}~\bibnamefont
  {Mazzoni}},\ }\href {https://doi.org/10.1007/JHEP12(2023)144} {\bibfield
  {journal} {\bibinfo  {journal} {J. High Energy Phys.}\ }\textbf {\bibinfo
  {volume} {2023}}\bibinfo  {number} { (12)},\ \bibinfo {pages}
  {144}}\BibitemShut {NoStop}%
\bibitem [{\citenamefont {Capizzi}\ and\ \citenamefont
  {Vitale}(2023)}]{capizzi2023universal}%
  \BibitemOpen
\bibfield  {number} {  }\bibfield  {author} {\bibinfo {author} {\bibfnamefont
  {L.}~\bibnamefont {Capizzi}}\ and\ \bibinfo {author} {\bibfnamefont
  {V.}~\bibnamefont {Vitale}},\ }\href {https://arxiv.org/abs/2310.01962}
  {\bibfield  {journal} {\bibinfo  {journal} {arXiv:2310.01962}\ } (\bibinfo
  {year} {2023})}\BibitemShut {NoStop}%
\bibitem [{\citenamefont {Ares}\ \emph {et~al.}(2024)\citenamefont {Ares},
  \citenamefont {Murciano}, \citenamefont {Piroli},\ and\ \citenamefont
  {Calabrese}}]{ares20233}%
  \BibitemOpen
  \bibfield  {author} {\bibinfo {author} {\bibfnamefont {F.}~\bibnamefont
  {Ares}}, \bibinfo {author} {\bibfnamefont {S.}~\bibnamefont {Murciano}},
  \bibinfo {author} {\bibfnamefont {L.}~\bibnamefont {Piroli}},\ and\ \bibinfo
  {author} {\bibfnamefont {P.}~\bibnamefont {Calabrese}},\ }\href
  {https://doi.org/10.1103/PhysRevD.110.L061901} {\bibfield  {journal}
  {\bibinfo  {journal} {Phys. Rev. D}\ }\textbf {\bibinfo {volume} {110}},\
  \bibinfo {pages} {L061901} (\bibinfo {year} {2024})}\BibitemShut {NoStop}%
\bibitem [{\citenamefont {Khor}\ \emph {et~al.}(2024)\citenamefont {Khor},
  \citenamefont {K{\"{u}}rk{\c{c}}{\"{u}}oglu}, \citenamefont {Hobbs},
  \citenamefont {Perdue},\ and\ \citenamefont {Klich}}]{khor2023}%
  \BibitemOpen
  \bibfield  {author} {\bibinfo {author} {\bibfnamefont {B.~J.~J.}\
  \bibnamefont {Khor}}, \bibinfo {author} {\bibfnamefont {D.~M.}\ \bibnamefont
  {K{\"{u}}rk{\c{c}}{\"{u}}oglu}}, \bibinfo {author} {\bibfnamefont {T.~J.}\
  \bibnamefont {Hobbs}}, \bibinfo {author} {\bibfnamefont {G.~N.}\ \bibnamefont
  {Perdue}},\ and\ \bibinfo {author} {\bibfnamefont {I.}~\bibnamefont
  {Klich}},\ }\href {https://doi.org/10.22331/q-2024-09-06-1462} {\bibfield
  {journal} {\bibinfo  {journal} {{Quantum}}\ }\textbf {\bibinfo {volume}
  {8}},\ \bibinfo {pages} {1462} (\bibinfo {year} {2024})}\BibitemShut
  {NoStop}%
\bibitem [{\citenamefont {Ma}\ \emph {et~al.}(2022)\citenamefont {Ma},
  \citenamefont {Han}, \citenamefont {Meir},\ and\ \citenamefont
  {Sela}}]{Ma2022}%
  \BibitemOpen
  \bibfield  {author} {\bibinfo {author} {\bibfnamefont {Z.}~\bibnamefont
  {Ma}}, \bibinfo {author} {\bibfnamefont {C.}~\bibnamefont {Han}}, \bibinfo
  {author} {\bibfnamefont {Y.}~\bibnamefont {Meir}},\ and\ \bibinfo {author}
  {\bibfnamefont {E.}~\bibnamefont {Sela}},\ }\href
  {https://doi.org/10.1103/PhysRevA.105.042416} {\bibfield  {journal} {\bibinfo
   {journal} {Phys. Rev. A}\ }\textbf {\bibinfo {volume} {105}},\ \bibinfo
  {pages} {042416} (\bibinfo {year} {2022})}\BibitemShut {NoStop}%
\bibitem [{\citenamefont {Han}\ \emph {et~al.}(2023)\citenamefont {Han},
  \citenamefont {Meir},\ and\ \citenamefont {Sela}}]{Han2023}%
  \BibitemOpen
  \bibfield  {author} {\bibinfo {author} {\bibfnamefont {C.}~\bibnamefont
  {Han}}, \bibinfo {author} {\bibfnamefont {Y.}~\bibnamefont {Meir}},\ and\
  \bibinfo {author} {\bibfnamefont {E.}~\bibnamefont {Sela}},\ }\href
  {https://doi.org/10.1103/PhysRevLett.130.136201} {\bibfield  {journal}
  {\bibinfo  {journal} {Phys. Rev. Lett.}\ }\textbf {\bibinfo {volume} {130}},\
  \bibinfo {pages} {136201} (\bibinfo {year} {2023})}\BibitemShut {NoStop}%
\bibitem [{\citenamefont {{Supplemental material.}}()}]{supmat}%
  \BibitemOpen
  \bibfield  {author} {\bibinfo {author} {\bibnamefont {{Supplemental
  material.}}},\ }\href@noop {} {}\BibitemShut {NoStop}%
\bibitem [{\citenamefont {Derrida}\ and\ \citenamefont
  {Gerschenfeld}(2009)}]{Derrida2009}%
  \BibitemOpen
  \bibfield  {author} {\bibinfo {author} {\bibfnamefont {B.}~\bibnamefont
  {Derrida}}\ and\ \bibinfo {author} {\bibfnamefont {A.}~\bibnamefont
  {Gerschenfeld}},\ }\href {https://doi.org/10.1007/s10955-009-9830-1}
  {\bibfield  {journal} {\bibinfo  {journal} {J Stat. Phys.}\ }\textbf
  {\bibinfo {volume} {137}},\ \bibinfo {pages} {978} (\bibinfo {year}
  {2009})}\BibitemShut {NoStop}%
\bibitem [{\citenamefont {Saha}\ and\ \citenamefont {Sadhu}(2023)}]{Saha_2023}%
  \BibitemOpen
  \bibfield  {author} {\bibinfo {author} {\bibfnamefont {S.}~\bibnamefont
  {Saha}}\ and\ \bibinfo {author} {\bibfnamefont {T.}~\bibnamefont {Sadhu}},\
  }\href {https://doi.org/10.1088/1742-5468/ace3b2} {\bibfield  {journal}
  {\bibinfo  {journal} {J. Stat. Mech.}\ }\textbf {\bibinfo {volume} {2023}},\
  \bibinfo {pages} {073207} (\bibinfo {year} {2023})}\BibitemShut {NoStop}%
\bibitem [{\citenamefont {Potter}\ and\ \citenamefont
  {Vasseur}(2022)}]{Potter2022}%
  \BibitemOpen
  \bibfield  {author} {\bibinfo {author} {\bibfnamefont {A.~C.}\ \bibnamefont
  {Potter}}\ and\ \bibinfo {author} {\bibfnamefont {R.}~\bibnamefont
  {Vasseur}},\ }\bibinfo {title} {Entanglement dynamics in hybrid quantum
  circuits},\ in\ \href {https://doi.org/10.1007/978-3-031-03998-0_9} {\emph
  {\bibinfo {booktitle} {Entanglement in Spin Chains}}}\ (\bibinfo  {publisher}
  {Springer International Publishing},\ \bibinfo {year} {2022})\ p.\ \bibinfo
  {pages} {211–249}\BibitemShut {NoStop}%
\bibitem [{\citenamefont {Skinner}\ \emph {et~al.}(2019)\citenamefont
  {Skinner}, \citenamefont {Ruhman},\ and\ \citenamefont
  {Nahum}}]{skinner2019}%
  \BibitemOpen
  \bibfield  {author} {\bibinfo {author} {\bibfnamefont {B.}~\bibnamefont
  {Skinner}}, \bibinfo {author} {\bibfnamefont {J.}~\bibnamefont {Ruhman}},\
  and\ \bibinfo {author} {\bibfnamefont {A.}~\bibnamefont {Nahum}},\ }\href
  {https://doi.org/10.1103/PhysRevX.9.031009} {\bibfield  {journal} {\bibinfo
  {journal} {Phys. Rev. X}\ }\textbf {\bibinfo {volume} {9}},\ \bibinfo {pages}
  {031009} (\bibinfo {year} {2019})}\BibitemShut {NoStop}%
\bibitem [{\citenamefont {Sierant}\ \emph {et~al.}(2022)\citenamefont
  {Sierant}, \citenamefont {Schir\`o}, \citenamefont {Lewenstein},\ and\
  \citenamefont {Turkeshi}}]{sierant2022}%
  \BibitemOpen
  \bibfield  {author} {\bibinfo {author} {\bibfnamefont {P.}~\bibnamefont
  {Sierant}}, \bibinfo {author} {\bibfnamefont {M.}~\bibnamefont {Schir\`o}},
  \bibinfo {author} {\bibfnamefont {M.}~\bibnamefont {Lewenstein}},\ and\
  \bibinfo {author} {\bibfnamefont {X.}~\bibnamefont {Turkeshi}},\ }\href
  {https://doi.org/10.1103/PhysRevB.106.214316} {\bibfield  {journal} {\bibinfo
   {journal} {Phys. Rev. B}\ }\textbf {\bibinfo {volume} {106}},\ \bibinfo
  {pages} {214316} (\bibinfo {year} {2022})}\BibitemShut {NoStop}%
\bibitem [{\citenamefont {Sierant}\ and\ \citenamefont
  {Turkeshi}(2022)}]{sierant20222}%
  \BibitemOpen
  \bibfield  {author} {\bibinfo {author} {\bibfnamefont {P.}~\bibnamefont
  {Sierant}}\ and\ \bibinfo {author} {\bibfnamefont {X.}~\bibnamefont
  {Turkeshi}},\ }\href {https://doi.org/10.1103/PhysRevLett.128.130605}
  {\bibfield  {journal} {\bibinfo  {journal} {Phys. Rev. Lett.}\ }\textbf
  {\bibinfo {volume} {128}},\ \bibinfo {pages} {130605} (\bibinfo {year}
  {2022})}\BibitemShut {NoStop}%
\bibitem [{\citenamefont {Agrawal}\ \emph {et~al.}(2022)\citenamefont
  {Agrawal}, \citenamefont {Zabalo}, \citenamefont {Chen}, \citenamefont
  {Wilson}, \citenamefont {Potter}, \citenamefont {Pixley}, \citenamefont
  {Gopalakrishnan},\ and\ \citenamefont {Vasseur}}]{agrawal2022}%
  \BibitemOpen
  \bibfield  {author} {\bibinfo {author} {\bibfnamefont {U.}~\bibnamefont
  {Agrawal}}, \bibinfo {author} {\bibfnamefont {A.}~\bibnamefont {Zabalo}},
  \bibinfo {author} {\bibfnamefont {K.}~\bibnamefont {Chen}}, \bibinfo {author}
  {\bibfnamefont {J.~H.}\ \bibnamefont {Wilson}}, \bibinfo {author}
  {\bibfnamefont {A.~C.}\ \bibnamefont {Potter}}, \bibinfo {author}
  {\bibfnamefont {J.~H.}\ \bibnamefont {Pixley}}, \bibinfo {author}
  {\bibfnamefont {S.}~\bibnamefont {Gopalakrishnan}},\ and\ \bibinfo {author}
  {\bibfnamefont {R.}~\bibnamefont {Vasseur}},\ }\href
  {https://doi.org/10.1103/PhysRevX.12.041002} {\bibfield  {journal} {\bibinfo
  {journal} {Phys. Rev. X}\ }\textbf {\bibinfo {volume} {12}},\ \bibinfo
  {pages} {041002} (\bibinfo {year} {2022})}\BibitemShut {NoStop}%
\bibitem [{\citenamefont {Barratt}\ \emph {et~al.}(2022)\citenamefont
  {Barratt}, \citenamefont {Agrawal}, \citenamefont {Gopalakrishnan},
  \citenamefont {Huse}, \citenamefont {Vasseur},\ and\ \citenamefont
  {Potter}}]{barratt2022}%
  \BibitemOpen
  \bibfield  {author} {\bibinfo {author} {\bibfnamefont {F.}~\bibnamefont
  {Barratt}}, \bibinfo {author} {\bibfnamefont {U.}~\bibnamefont {Agrawal}},
  \bibinfo {author} {\bibfnamefont {S.}~\bibnamefont {Gopalakrishnan}},
  \bibinfo {author} {\bibfnamefont {D.~A.}\ \bibnamefont {Huse}}, \bibinfo
  {author} {\bibfnamefont {R.}~\bibnamefont {Vasseur}},\ and\ \bibinfo {author}
  {\bibfnamefont {A.~C.}\ \bibnamefont {Potter}},\ }\href
  {https://doi.org/10.1103/PhysRevLett.129.120604} {\bibfield  {journal}
  {\bibinfo  {journal} {Phys. Rev. Lett.}\ }\textbf {\bibinfo {volume} {129}},\
  \bibinfo {pages} {120604} (\bibinfo {year} {2022})}\BibitemShut {NoStop}%
\bibitem [{\citenamefont {Oshima}\ and\ \citenamefont
  {Fuji}(2023)}]{oshima2023}%
  \BibitemOpen
  \bibfield  {author} {\bibinfo {author} {\bibfnamefont {H.}~\bibnamefont
  {Oshima}}\ and\ \bibinfo {author} {\bibfnamefont {Y.}~\bibnamefont {Fuji}},\
  }\href {https://doi.org/10.1103/PhysRevB.107.014308} {\bibfield  {journal}
  {\bibinfo  {journal} {Phys. Rev. B}\ }\textbf {\bibinfo {volume} {107}},\
  \bibinfo {pages} {014308} (\bibinfo {year} {2023})}\BibitemShut {NoStop}%
\bibitem [{\citenamefont {Klocke}\ and\ \citenamefont
  {Buchhold}(2023)}]{klocke2023}%
  \BibitemOpen
  \bibfield  {author} {\bibinfo {author} {\bibfnamefont {K.}~\bibnamefont
  {Klocke}}\ and\ \bibinfo {author} {\bibfnamefont {M.}~\bibnamefont
  {Buchhold}},\ }\href {https://doi.org/10.1103/PhysRevX.13.041028} {\bibfield
  {journal} {\bibinfo  {journal} {Phys. Rev. X}\ }\textbf {\bibinfo {volume}
  {13}},\ \bibinfo {pages} {041028} (\bibinfo {year} {2023})}\BibitemShut
  {NoStop}%
\bibitem [{\citenamefont {Cao}\ \emph {et~al.}(2019)\citenamefont {Cao},
  \citenamefont {Tilloy},\ and\ \citenamefont {Luca}}]{cao2019}%
  \BibitemOpen
  \bibfield  {author} {\bibinfo {author} {\bibfnamefont {X.}~\bibnamefont
  {Cao}}, \bibinfo {author} {\bibfnamefont {A.}~\bibnamefont {Tilloy}},\ and\
  \bibinfo {author} {\bibfnamefont {A.~D.}\ \bibnamefont {Luca}},\ }\href
  {https://doi.org/10.21468/SciPostPhys.7.2.024} {\bibfield  {journal}
  {\bibinfo  {journal} {SciPost Phys.}\ }\textbf {\bibinfo {volume} {7}},\
  \bibinfo {pages} {024} (\bibinfo {year} {2019})}\BibitemShut {NoStop}%
\bibitem [{\citenamefont {Alberton}\ \emph {et~al.}(2021)\citenamefont
  {Alberton}, \citenamefont {Buchhold},\ and\ \citenamefont
  {Diehl}}]{alberton2021}%
  \BibitemOpen
  \bibfield  {author} {\bibinfo {author} {\bibfnamefont {O.}~\bibnamefont
  {Alberton}}, \bibinfo {author} {\bibfnamefont {M.}~\bibnamefont {Buchhold}},\
  and\ \bibinfo {author} {\bibfnamefont {S.}~\bibnamefont {Diehl}},\ }\href
  {https://doi.org/10.1103/PhysRevLett.126.170602} {\bibfield  {journal}
  {\bibinfo  {journal} {Phys. Rev. Lett.}\ }\textbf {\bibinfo {volume} {126}},\
  \bibinfo {pages} {170602} (\bibinfo {year} {2021})}\BibitemShut {NoStop}%
\bibitem [{\citenamefont {Buchhold}\ \emph {et~al.}(2021)\citenamefont
  {Buchhold}, \citenamefont {Minoguchi}, \citenamefont {Altland},\ and\
  \citenamefont {Diehl}}]{buchhold2021}%
  \BibitemOpen
  \bibfield  {author} {\bibinfo {author} {\bibfnamefont {M.}~\bibnamefont
  {Buchhold}}, \bibinfo {author} {\bibfnamefont {Y.}~\bibnamefont {Minoguchi}},
  \bibinfo {author} {\bibfnamefont {A.}~\bibnamefont {Altland}},\ and\ \bibinfo
  {author} {\bibfnamefont {S.}~\bibnamefont {Diehl}},\ }\href
  {https://doi.org/10.1103/PhysRevX.11.041004} {\bibfield  {journal} {\bibinfo
  {journal} {Phys. Rev. X}\ }\textbf {\bibinfo {volume} {11}},\ \bibinfo
  {pages} {041004} (\bibinfo {year} {2021})}\BibitemShut {NoStop}%
\bibitem [{\citenamefont {Fava}\ \emph {et~al.}(2023)\citenamefont {Fava},
  \citenamefont {Piroli}, \citenamefont {Swann}, \citenamefont {Bernard},\ and\
  \citenamefont {Nahum}}]{fava2023}%
  \BibitemOpen
  \bibfield  {author} {\bibinfo {author} {\bibfnamefont {M.}~\bibnamefont
  {Fava}}, \bibinfo {author} {\bibfnamefont {L.}~\bibnamefont {Piroli}},
  \bibinfo {author} {\bibfnamefont {T.}~\bibnamefont {Swann}}, \bibinfo
  {author} {\bibfnamefont {D.}~\bibnamefont {Bernard}},\ and\ \bibinfo {author}
  {\bibfnamefont {A.}~\bibnamefont {Nahum}},\ }\href
  {https://doi.org/10.1103/PhysRevX.13.041045} {\bibfield  {journal} {\bibinfo
  {journal} {Phys. Rev. X}\ }\textbf {\bibinfo {volume} {13}},\ \bibinfo
  {pages} {041045} (\bibinfo {year} {2023})}\BibitemShut {NoStop}%
\bibitem [{\citenamefont {Poboiko}\ \emph {et~al.}(2023)\citenamefont
  {Poboiko}, \citenamefont {P\"opperl}, \citenamefont {Gornyi},\ and\
  \citenamefont {Mirlin}}]{poboiko2023}%
  \BibitemOpen
  \bibfield  {author} {\bibinfo {author} {\bibfnamefont {I.}~\bibnamefont
  {Poboiko}}, \bibinfo {author} {\bibfnamefont {P.}~\bibnamefont {P\"opperl}},
  \bibinfo {author} {\bibfnamefont {I.~V.}\ \bibnamefont {Gornyi}},\ and\
  \bibinfo {author} {\bibfnamefont {A.~D.}\ \bibnamefont {Mirlin}},\ }\href
  {https://doi.org/10.1103/PhysRevX.13.041046} {\bibfield  {journal} {\bibinfo
  {journal} {Phys. Rev. X}\ }\textbf {\bibinfo {volume} {13}},\ \bibinfo
  {pages} {041046} (\bibinfo {year} {2023})}\BibitemShut {NoStop}%
\bibitem [{\citenamefont {M\"uller}\ \emph {et~al.}(2022)\citenamefont
  {M\"uller}, \citenamefont {Diehl},\ and\ \citenamefont
  {Buchhold}}]{muller2022}%
  \BibitemOpen
  \bibfield  {author} {\bibinfo {author} {\bibfnamefont {T.}~\bibnamefont
  {M\"uller}}, \bibinfo {author} {\bibfnamefont {S.}~\bibnamefont {Diehl}},\
  and\ \bibinfo {author} {\bibfnamefont {M.}~\bibnamefont {Buchhold}},\ }\href
  {https://doi.org/10.1103/PhysRevLett.128.010605} {\bibfield  {journal}
  {\bibinfo  {journal} {Phys. Rev. Lett.}\ }\textbf {\bibinfo {volume} {128}},\
  \bibinfo {pages} {010605} (\bibinfo {year} {2022})}\BibitemShut {NoStop}%
\bibitem [{\citenamefont {L\'oio}\ \emph {et~al.}(2023)\citenamefont {L\'oio},
  \citenamefont {De~Luca}, \citenamefont {De~Nardis},\ and\ \citenamefont
  {Turkeshi}}]{loio2023}%
  \BibitemOpen
  \bibfield  {author} {\bibinfo {author} {\bibfnamefont {H.}~\bibnamefont
  {L\'oio}}, \bibinfo {author} {\bibfnamefont {A.}~\bibnamefont {De~Luca}},
  \bibinfo {author} {\bibfnamefont {J.}~\bibnamefont {De~Nardis}},\ and\
  \bibinfo {author} {\bibfnamefont {X.}~\bibnamefont {Turkeshi}},\ }\href
  {https://doi.org/10.1103/PhysRevB.108.L020306} {\bibfield  {journal}
  {\bibinfo  {journal} {Phys. Rev. B}\ }\textbf {\bibinfo {volume} {108}},\
  \bibinfo {pages} {L020306} (\bibinfo {year} {2023})}\BibitemShut {NoStop}%
\bibitem [{\citenamefont {Poboiko}\ \emph {et~al.}(2024)\citenamefont
  {Poboiko}, \citenamefont {Gornyi},\ and\ \citenamefont
  {Mirlin}}]{poboiko2023measurementinduced}%
  \BibitemOpen
  \bibfield  {author} {\bibinfo {author} {\bibfnamefont {I.}~\bibnamefont
  {Poboiko}}, \bibinfo {author} {\bibfnamefont {I.~V.}\ \bibnamefont
  {Gornyi}},\ and\ \bibinfo {author} {\bibfnamefont {A.~D.}\ \bibnamefont
  {Mirlin}},\ }\href {https://doi.org/10.1103/PhysRevLett.132.110403}
  {\bibfield  {journal} {\bibinfo  {journal} {Phys. Rev. Lett.}\ }\textbf
  {\bibinfo {volume} {132}},\ \bibinfo {pages} {110403} (\bibinfo {year}
  {2024})}\BibitemShut {NoStop}%
\bibitem [{\citenamefont {Chahine}\ and\ \citenamefont
  {Buchhold}(2024)}]{chahine2023entanglement}%
  \BibitemOpen
  \bibfield  {author} {\bibinfo {author} {\bibfnamefont {K.}~\bibnamefont
  {Chahine}}\ and\ \bibinfo {author} {\bibfnamefont {M.}~\bibnamefont
  {Buchhold}},\ }\href {https://doi.org/10.1103/PhysRevB.110.054313} {\bibfield
   {journal} {\bibinfo  {journal} {Phys. Rev. B}\ }\textbf {\bibinfo {volume}
  {110}},\ \bibinfo {pages} {054313} (\bibinfo {year} {2024})}\BibitemShut
  {NoStop}%
\bibitem [{\citenamefont {Turkeshi}\ \emph {et~al.}(2022)\citenamefont
  {Turkeshi}, \citenamefont {Piroli},\ and\ \citenamefont
  {Schir\'o}}]{turkeshi2022e}%
  \BibitemOpen
  \bibfield  {author} {\bibinfo {author} {\bibfnamefont {X.}~\bibnamefont
  {Turkeshi}}, \bibinfo {author} {\bibfnamefont {L.}~\bibnamefont {Piroli}},\
  and\ \bibinfo {author} {\bibfnamefont {M.}~\bibnamefont {Schir\'o}},\ }\href
  {https://doi.org/10.1103/PhysRevB.106.024304} {\bibfield  {journal} {\bibinfo
   {journal} {Phys. Rev. B}\ }\textbf {\bibinfo {volume} {106}},\ \bibinfo
  {pages} {024304} (\bibinfo {year} {2022})}\BibitemShut {NoStop}%
\bibitem [{\citenamefont {Turkeshi}\ \emph {et~al.}(2024)\citenamefont
  {Turkeshi}, \citenamefont {Piroli},\ and\ \citenamefont
  {Schir\`o}}]{turkeshi2024density}%
  \BibitemOpen
  \bibfield  {author} {\bibinfo {author} {\bibfnamefont {X.}~\bibnamefont
  {Turkeshi}}, \bibinfo {author} {\bibfnamefont {L.}~\bibnamefont {Piroli}},\
  and\ \bibinfo {author} {\bibfnamefont {M.}~\bibnamefont {Schir\`o}},\ }\href
  {https://doi.org/10.1103/PhysRevB.109.144306} {\bibfield  {journal} {\bibinfo
   {journal} {Phys. Rev. B}\ }\textbf {\bibinfo {volume} {109}},\ \bibinfo
  {pages} {144306} (\bibinfo {year} {2024})}\BibitemShut {NoStop}%
\bibitem [{\citenamefont {Gal}\ \emph {et~al.}(2023)\citenamefont {Gal},
  \citenamefont {Turkeshi},\ and\ \citenamefont {Schirò}}]{legal2023}%
  \BibitemOpen
  \bibfield  {author} {\bibinfo {author} {\bibfnamefont {Y.~L.}\ \bibnamefont
  {Gal}}, \bibinfo {author} {\bibfnamefont {X.}~\bibnamefont {Turkeshi}},\ and\
  \bibinfo {author} {\bibfnamefont {M.}~\bibnamefont {Schirò}},\ }\href
  {https://doi.org/10.21468/SciPostPhys.14.5.138} {\bibfield  {journal}
  {\bibinfo  {journal} {SciPost Phys.}\ }\textbf {\bibinfo {volume} {14}},\
  \bibinfo {pages} {138} (\bibinfo {year} {2023})}\BibitemShut {NoStop}%
\bibitem [{\citenamefont {Le~Gal}\ \emph {et~al.}(2024)\citenamefont {Le~Gal},
  \citenamefont {Turkeshi},\ and\ \citenamefont
  {Schir\`o}}]{gal2024entanglement}%
  \BibitemOpen
  \bibfield  {author} {\bibinfo {author} {\bibfnamefont {Y.}~\bibnamefont
  {Le~Gal}}, \bibinfo {author} {\bibfnamefont {X.}~\bibnamefont {Turkeshi}},\
  and\ \bibinfo {author} {\bibfnamefont {M.}~\bibnamefont {Schir\`o}},\ }\href
  {https://doi.org/10.1103/PRXQuantum.5.030329} {\bibfield  {journal} {\bibinfo
   {journal} {PRX Quantum}\ }\textbf {\bibinfo {volume} {5}},\ \bibinfo {pages}
  {030329} (\bibinfo {year} {2024})}\BibitemShut {NoStop}%
\bibitem [{\citenamefont {Leung}\ \emph {et~al.}()\citenamefont {Leung},
  \citenamefont {Meidan},\ and\ \citenamefont {Romito}}]{leung2023theory}%
  \BibitemOpen
  \bibfield  {author} {\bibinfo {author} {\bibfnamefont {C.~Y.}\ \bibnamefont
  {Leung}}, \bibinfo {author} {\bibfnamefont {D.}~\bibnamefont {Meidan}},\ and\
  \bibinfo {author} {\bibfnamefont {A.}~\bibnamefont {Romito}},\ }\href@noop {}
  {}\Eprint {https://arxiv.org/abs/2312.14022} {arXiv:2312.14022} \BibitemShut
  {NoStop}%
\bibitem [{\citenamefont {{R.-A. Chang and M. Ippoliti}}(2024)}]{ippo}%
  \BibitemOpen
  \bibfield  {author} {\bibinfo {author} {\bibnamefont {{R.-A. Chang and M.
  Ippoliti}}},\ }\href@noop {} {\bibfield  {journal} {\bibinfo  {journal} {To
  appear}\ } (\bibinfo {year} {2024})}\BibitemShut {NoStop}%
\bibitem [{\citenamefont {Turkeshi}\ \emph {et~al.}()\citenamefont {Turkeshi},
  \citenamefont {{De Luca}},\ and\ \citenamefont {Calabrese}}]{dataavail}%
  \BibitemOpen
  \bibfield  {author} {\bibinfo {author} {\bibfnamefont {X.}~\bibnamefont
  {Turkeshi}}, \bibinfo {author} {\bibfnamefont {A.}~\bibnamefont {{De
  Luca}}},\ and\ \bibinfo {author} {\bibfnamefont {P.}~\bibnamefont
  {Calabrese}},\ }\href@noop {} {\bibinfo {title} {Code available in zenodo for
  publication.}}\BibitemShut {Stop}%
\bibitem [{\citenamefont {Liu}\ \emph {et~al.}(2024)\citenamefont {Liu},
  \citenamefont {Zhang}, \citenamefont {Yin},\ and\ \citenamefont
  {Zhang}}]{liu2024symmetry}%
  \BibitemOpen
  \bibfield  {author} {\bibinfo {author} {\bibfnamefont {S.}~\bibnamefont
  {Liu}}, \bibinfo {author} {\bibfnamefont {H.-K.}\ \bibnamefont {Zhang}},
  \bibinfo {author} {\bibfnamefont {S.}~\bibnamefont {Yin}},\ and\ \bibinfo
  {author} {\bibfnamefont {S.-X.}\ \bibnamefont {Zhang}},\ }\href
  {https://doi.org/10.1103/PhysRevLett.133.140405} {\bibfield  {journal}
  {\bibinfo  {journal} {Phys. Rev. Lett.}\ }\textbf {\bibinfo {volume} {133}},\
  \bibinfo {pages} {140405} (\bibinfo {year} {2024})}\BibitemShut {NoStop}%
\bibitem [{Note3()}]{Note3}%
  \BibitemOpen
  \bibinfo {note} {The replica limit $n\to 1$ is complicated by a non-trivial
  interplay between hydrodynamic and non-hydrodynamic modes~\cite
  {keyserlingk2018}.}\BibitemShut {Stop}%
\bibitem [{\citenamefont {Collins}\ and\ \citenamefont
  {Śniady}(2006)}]{Collins2006}%
  \BibitemOpen
  \bibfield  {author} {\bibinfo {author} {\bibfnamefont {B.}~\bibnamefont
  {Collins}}\ and\ \bibinfo {author} {\bibfnamefont {P.}~\bibnamefont
  {Śniady}},\ }\href {https://doi.org/10.1007/s00220-006-1554-3} {\bibfield
  {journal} {\bibinfo  {journal} {Commun. Math. Phys.}\ }\textbf {\bibinfo
  {volume} {264}},\ \bibinfo {pages} {773} (\bibinfo {year}
  {2006})}\BibitemShut {NoStop}%
\bibitem [{\citenamefont {Fishman}\ \emph {et~al.}(2022)\citenamefont
  {Fishman}, \citenamefont {White},\ and\ \citenamefont
  {Stoudenmire}}]{itensor}%
  \BibitemOpen
  \bibfield  {author} {\bibinfo {author} {\bibfnamefont {M.}~\bibnamefont
  {Fishman}}, \bibinfo {author} {\bibfnamefont {S.~R.}\ \bibnamefont {White}},\
  and\ \bibinfo {author} {\bibfnamefont {E.~M.}\ \bibnamefont {Stoudenmire}},\
  }\href {https://doi.org/10.21468/SciPostPhysCodeb.4} {\bibfield  {journal}
  {\bibinfo  {journal} {SciPost Phys. Codebases}\ ,\ \bibinfo {pages} {4}}
  (\bibinfo {year} {2022})}\BibitemShut {NoStop}%
\bibitem [{\citenamefont {Schollwöck}(2011)}]{Schollw_ck_2011}%
  \BibitemOpen
  \bibfield  {author} {\bibinfo {author} {\bibfnamefont {U.}~\bibnamefont
  {Schollwöck}},\ }\href {https://doi.org/10.1016/j.aop.2010.09.012}
  {\bibfield  {journal} {\bibinfo  {journal} {Ann. Phys.}\ }\textbf {\bibinfo
  {volume} {326}},\ \bibinfo {pages} {96} (\bibinfo {year} {2011})}\BibitemShut
  {NoStop}%
\bibitem [{\citenamefont {Turkeshi}\ and\ \citenamefont
  {Sierant}(2024)}]{turkeshi2024hilbert}%
  \BibitemOpen
  \bibfield  {author} {\bibinfo {author} {\bibfnamefont {X.}~\bibnamefont
  {Turkeshi}}\ and\ \bibinfo {author} {\bibfnamefont {P.}~\bibnamefont
  {Sierant}},\ }\href {https://doi.org/10.3390/e26060471} {\bibfield  {journal}
  {\bibinfo  {journal} {Entropy}\ }\textbf {\bibinfo {volume} {26}},\ \bibinfo
  {pages} {471} (\bibinfo {year} {2024})}\BibitemShut {NoStop}%
\bibitem [{\citenamefont {Khemani}\ \emph
  {et~al.}(2018{\natexlab{b}})\citenamefont {Khemani}, \citenamefont {Huse},\
  and\ \citenamefont {Nahum}}]{PhysRevB.98.144304}%
  \BibitemOpen
  \bibfield  {author} {\bibinfo {author} {\bibfnamefont {V.}~\bibnamefont
  {Khemani}}, \bibinfo {author} {\bibfnamefont {D.~A.}\ \bibnamefont {Huse}},\
  and\ \bibinfo {author} {\bibfnamefont {A.}~\bibnamefont {Nahum}},\ }\href
  {https://doi.org/10.1103/PhysRevB.98.144304} {\bibfield  {journal} {\bibinfo
  {journal} {Phys. Rev. B}\ }\textbf {\bibinfo {volume} {98}},\ \bibinfo
  {pages} {144304} (\bibinfo {year} {2018}{\natexlab{b}})}\BibitemShut
  {NoStop}%
\end{thebibliography}%
